\documentclass[12pt]{article}

\usepackage[dvipsname]{xcolor}

\usepackage{graphicx}
\usepackage{amsmath}        
\usepackage{amssymb}        
\usepackage{slashed}
\usepackage{bm}
\usepackage{here}
\usepackage{cite}

\def\mydate{May 31,  2022}
\def\ignore#1{{}}

\def\go{\rightarrow}
\def\dd{\partial}

\def\ep{{\epsilon}}

\def\KK{{\rm KK}}

\def\onehalf{\hbox{$\frac{1}{2}$}}
\def\la{\langle}
\def\ra{\rangle}
\def\flat{{\rm flat}}
\def\RS{{\rm RS}}
\def\Tr{{\rm Tr} \,}
\def\Psibar{\overline{\Psi}}
\def\diag{{\rm diag}\,}

\def\mybig{\displaystyle \strut }

\def\myfrac#1#2{\frac{\mybig #1}{\mybig #2}}

\def\mymat#1#2{\begin{matrix}#1 \cr \noalign{\kern -2pt} #2\end{matrix}}

\def\mynoalign{\noalign{\kern 4pt}}
\def\mysnoalign{\noalign{\kern 3pt}}

\makeatletter
\@addtoreset{equation}{section}
\makeatother

\begin{document}

\thispagestyle{empty}

{\small \noindent \mydate \hfill OU-HET-1144}

\vskip 3.5cm

\baselineskip=30pt plus 1pt minus 1pt

\begin{center}
{\LARGE \bf  Universality in Anomaly Flow}

\end{center}


\baselineskip=22pt plus 1pt minus 1pt

\vskip 1.5cm

\begin{center}
{\large \bf  Yutaka Hosotani}

\baselineskip=18pt plus 1pt minus 1pt

\vskip 10pt
{\small \it Department of Physics, Osaka University,  Toyonaka, Osaka 560-0043, Japan} \\

\end{center}

\vskip 3.cm
\baselineskip=18pt plus 1pt minus 1pt

\begin{abstract}
Universality in anomaly flow by an Aharonov-Bohm (AB) phase $\theta_H$ is shown in the flat 
$M^4 \times (S^1/Z_2)$  spacetime and in the Randall-Sundrum (RS) warped  space. 
We analyze $SU(2)$ gauge theory with doublet fermions. 
With orbifold boundary conditions the $U(1)$ part of gauge symmetry remains unbroken 
at $\theta_H = 0$ and $\pi$.
Chiral anomalies smoothly vary with $\theta_H$ in the RS space.
It is shown that  anomaly coefficients associated with this anomaly flow are expressed
in terms of the values of the wave functions of gauge fields at the  UV and IR branes in the RS space.
The anomaly coefficients depend on $\theta_H$, the warp factor of the RS space, and the orbifold boundary
conditions for  fermions, but not on the bulk mass parameters of  fermions.
\end{abstract}


\newpage

\baselineskip=20pt plus 1pt minus 1pt
\parskip=0pt

\section{Introduction} 

In gauge-Higgs unification (GHU), gauge symmetry is dynamically broken by an Aharonov-Bohm (AB) 
phase, $\theta_H$, in the fifth dimension\cite{Hosotani1983, Davies1988, Hosotani1989, Davies1989, 
Hatanaka1998, Hatanaka1999, Kubo2002}.
It has been shown recently that chiral anomalies \cite{Adler1969, BellJackiw1969, Fujikawa1979, Fujikawa1980} 
 in GHU flow with $\theta_H$, that is, anomaly coefficients 
smoothly change with $\theta_H$ in the Randall-Sundrum (RS) warped space \cite{anomalyFlow2022}.
In the GUT-inspired $SO(5) \times U(1)_X \times SU(3)_C$ GHU models 
in the RS space,  chiral quarks and leptons  at $\theta_H = 0$ are transformed 
to vector-like fermions at $\theta_H = \pi$\cite{FiniteT2021}. 
As $\theta_H$ varies from 0 to $\pi$, $SU(2)_L \times U(1)_Y \times SU(3)_C$ gauge symmetry is
converted to $SU(2)_R \times U(1)_{Y'} \times SU(3)_C$ gauge symmetry.
Chiral fermions appearing as zero modes of fermion multiplets in the spinor representation of $SO(5)$
at $\theta_H = 0$ become massive fermions having vector-like gauge couplings at $\theta_H = \pi$.
The chiral anomaly induced by each quark or lepton at $\theta_H =0$ smoothly changes and 
vanishes at $\theta_H = \pi$.

In the RS space each fermion multiplet is characterized by  its own dimensionless bulk mass parameter $c$ 
which controls the mass and wave function of the fermion.  
In the previous paper \cite{anomalyFlow2022}  it has been recognized by numerical evaluation 
that the anomaly coefficients  depend on $\theta_H$, but not on the bulk mass parameter $c$.
This fact leads to a puzzle.  How can  the $\theta_H$-dependence of the anomaly coefficients be
determined and expressed independently of the details of the fermion field?
This is the main theme addressed in this paper.
We are going to show that the anomaly coefficients at general $\theta_H$ are expressed in terms of 
the values of the wave functions of gauge fields at the UV and IR branes in the RS space.
The anomaly coefficients depend on $\theta_H$, the warp factor $z_L$ of the RS space, and boundary conditions of
the fermion field, but not on the bulk mass parameter $c$.   The universality of the anomaly flow is observed.

We stress that the universal behavior is highly nontrivial.  In GHU in the RS space gauge couplings 
of each fermion mode depend on $\theta_H$, $z_L$  and $c$.  To find the total anomaly coefficients one needs to
sum all contributions coming from triangle loop diagrams in which all possible Kaluza-Klein (KK) excited modes
of fermions are running.  The universality of the anomaly flow is established only when all contributions are
taken into account.  

The phenomenon of anomaly flow is different from that of anomaly inflow in which 
anomalies and fermion zero modes on defects such as strings and domain walls or on the boundary of
spacetime are intertwined and related to each other \cite{CallanHarvey1985, Fukaya2017, WittenYonekura2021}.
In orbifold gauge theory gauge couplings of fermion modes vary with the AB phase $\theta_H$ in the fifth
dimension, and anomalies also vary with $\theta_H$.  We are going to show that the $\theta_H$-dependence
of the anomalies is expressed by a holographic formula involving the values of the wave functions of gauge fields.

In this paper we analyze  $SU(2)$ GHU models in the flat $M^4 \times (S^1/Z_2)$ spacetime and 
in the RS warped space  with orbifold boundary conditions  which break $SU(2)$ to $U(1)$.  
The $U(1)$ gauge symmetry survives at $\theta_H = 0$ and $\pi$. 
Fermion doublet multiplets have zero modes at $\theta_H = 0$ or $\pi$, depending on their boundary
conditions.
Chiral anomalies appear in various combinations of Kaluza-Klein (KK) modes of  gauge fields.
In the flat $M^4 \times (S^1/Z_2)$ spacetime all 4D gauge couplings are determined analytically, but
the KK mass spectrum of gauge and fermion fields exhibit level crossings as $\theta_H$ varies.
In the RS space there occurs no level crossing in the spectrum, 
and all gauge couplings smoothly vary with $\theta_H$.
The flat spacetime limit of the RS space gives rise to singular behavior of the anomalies as  functions 
of $\theta_H$, reproducing the known result in the flat spacetime.

In Section 2 $SU(2)$ GHU models are introduced both in flat  $M^4 \times (S^1/Z_2)$ spacetime
and in the RS space.  In Section 3 chiral anomalies are evaluated and expressed in a simple form 
which involves the values of the wave functions of gauge fields at the UV and IR branes and
boundary conditions of fermion fields.  In Section 4 conditions for anomaly cancellation are derived.
Section 5 is devoted to a summary and discussions.

\section{$SU(2)$ GHU}

We consider $SU(2)$ GHU in the flat $M^4 \times (S^1/Z_2)$ spacetime with coordinate 
$x^M$ ($M=0,1,2,3,5$, $x^5 =y$) whose action is given by
\begin{align}
I_\flat &= \int d^4 x \int_0^{L} dy  \, {\cal L}_\flat ~, \cr
\noalign{\kern 5pt}
{\cal L}_\flat & = - \frac{1}{2} \Tr  F_{MN} F^{MN} 
+  \Psibar \gamma^M D_M \Psi   ~,
\label{flataction}
\end{align}
where ${\cal L}_\flat (x^\mu, y)= {\cal L}_\flat (x^\mu, y + 2L) = {\cal L}_\flat (x^\mu, -y) $.
Here $F_{MN} = \dd_M A_N - \dd_N A_N - i g_A [A_M , A_N]$, $A_M = \onehalf  \sum_{a=1}^3 A^a_M \tau^a$
where $\tau^a$'s are Pauli matrices. 
We adopt the metric  $\eta_{MN}  = \diag (-1, 1, 1, 1, 1)$.
$\Psi$ is an $SU(2)$ doublet and $D_M = \dd_M - i g_A A_M$.
$\Psibar = i \Psi^\dagger \gamma^0$.
Orbifold boundary conditions are given, with $(y_0, y_1) = (0, L)$, by
\begin{align}
\begin{pmatrix} A_\mu \cr A_y \end{pmatrix} (x, y_j - y) 
&= P_j \begin{pmatrix} A_\mu \cr - A_y \end{pmatrix} (x,  y_j + y) P_j^{-1} ~, \cr
\noalign{\kern 5pt}
\Psi  (x, y_j - y)  &= 
\begin{cases} 
+ P_j \gamma^5 \Psi  (x, y_j + y) &\hbox{type 1A} \cr
\noalign{\kern 3pt}
-  P_j \gamma^5 \Psi  (x, y_j + y) &\hbox{type 1B} \cr
\noalign{\kern 3pt}
(-1)^j P_j  \gamma^5  \Psi  (x, y_j + y) &\hbox{type 2A} \cr
\noalign{\kern 3pt}
(-1)^{j+1} P_j  \gamma^5  \Psi  (x, y_j + y) &\hbox{type 2B} 
 \end{cases} , \cr
\noalign{\kern 5pt}
P_0 = P_1 &= \tau^3 ~.
\label{BC1}
\end{align}
The $SU(2)$ symmetry is broken to $U(1)$ by the boundary conditions (\ref{BC1}).
$A^3_\mu, A^{1,2}_y$ are parity even at both $y_0$ and $y_1$, and have constant zero modes.
The zero mode of $A^3_\mu$ is the 4D $U(1)$ gauge field,  and the 4D gauge coupling is given by
\begin{align}
g_4 = \frac{g_A}{\sqrt{L}} ~.
\label{coupling1}
\end{align}
We denote the doublet field  as $\Psi = (u, d)^t$.
In type 1A (1B) $u_R$ and $d_L$ ($u_L$ and $d_R$) are parity even  at both $y_0$ and $y_1$, and have zero modes, 
leading to chiral structure.

The zero modes of  $A_y^{1,2}$ may develop  nonvanishing expectation values.
Without loss of generality one may assume that $\la A_y^{1} \ra = 0$. 
An AB phase $\theta_H$ along the fifth dimension is given by
\begin{align}
&P \exp \bigg\{  i g_A \int_0^{2 L} dy \, \la A_y \ra \bigg\} = e^{ i \theta_H \tau^2 }
= \begin{pmatrix} \cos \theta_H & \sin\theta_H \cr - \sin\theta_H & \cos \theta_H \end{pmatrix}  ~, \cr
\noalign{\kern 5pt}
&\theta_H = g_4 L \, \la  A_y^{2} \ra ~.
\label{ABphase1}
\end{align}
The AB phase $\theta_H$ is a physical quantity.  It couples to fields, affecting their mass spectrum.
One can change the value of $\theta_H$ by a gauge transformation, which also alters
boundary conditions.  Under a large gauge transformation given by
\begin{align}
\tilde A_M &= \Omega A_M \Omega^{-1}  + \frac{i}{g_A} \Omega \dd_M \Omega^{-1} ~, ~~
\tilde \Psi = \Omega \Psi ~, \cr
\noalign{\kern 5pt}
\Omega &= \exp \Big( \frac{i}{2} \theta (y)  \tau^2 \Big) ~, ~~
\theta (y)  = \theta_H \Big( 1 - \frac{y}{L} \Big) ~, 
\label{ABphase2}
\end{align}
$\tilde \theta_H = 0$ and boundary condition matrices become
\begin{align}
\tilde P_j &= \Omega( y_j -y) P_j \Omega^{-1} (y_j + y) ~, \cr
\noalign{\kern 5pt}
\tilde P_0 &= \begin{pmatrix} \cos \theta_H & - \sin \theta_H \cr - \sin \theta_H & - \cos \theta_H \end{pmatrix} , ~~
\tilde P_1 = \tau^3 ~.
\label{BC2}
\end{align}
Although the AB phase $\tilde \theta_H$ vanishes, boundary conditions become nontrivial.  
Physics remains the same.  This gauge is called the twisted gauge\cite{Falkowski2007, HS2007}.

Fields in the twisted gauge satisfy free equations. 
KK expansions for $\tilde A_{\mu}^{1}, \, \tilde A_{\mu}^{3}$ are given by
\begin{align}
\begin{pmatrix} \tilde A_{\mu}^{1} (x,y) \cr \noalign{\kern 5pt} \tilde A_{\mu}^{3} (x,y) \end{pmatrix} 
&= \sum_{n=-\infty}^{\infty } B_{\mu}^{(n)} (x) 
\frac{1}{\sqrt{\pi R}}
\begin{pmatrix} \sin \Big[ \myfrac{ny}{R} - \theta (y) \Big] \cr \cos \Big[ \myfrac{ny}{R} - \theta (y) \Big] \end{pmatrix} 
\label{gaugeKKflat1}
\end{align}
where $L=\pi R$.
In the original gauge they become
\begin{align}
\begin{pmatrix} A_{\mu}^{1} (x,y) \cr \noalign{\kern 5pt}  A_{\mu}^{3} (x,y) \end{pmatrix} 
&= \sum_{n=-\infty}^{\infty } B_{\mu}^{(n )}  (x) 
\frac{1}{\sqrt{\pi R}}
\begin{pmatrix} \sin  \myfrac{ny}{R} \cr \cos  \myfrac{ny}{R}\end{pmatrix} .
\label{gaugeKKflat2}
\end{align}
The mass of the $B_{\mu}^{(n)}  (x)$ mode is 
$m_{n} (\theta_{H}) =R^{-1}\big| n + \frac{\theta_{H}}{\pi} \big|$.  
The spectrum is periodic in $\theta_H$ with period $\pi$.

Similarly the fermion field $\Psi$ in the twisted gauge 
\begin{align}
\tilde \Psi = \begin{pmatrix} \tilde u \cr \tilde d \end{pmatrix} 
= \begin{pmatrix} \cos \onehalf \theta(y) & \sin \onehalf \theta(y) \cr
- \sin \onehalf \theta(y) &  \cos \onehalf \theta(y) \end{pmatrix}
\begin{pmatrix} u \cr d \end{pmatrix}
\label{flatfermiwave1}
\end{align}
satisfies  free equations in the bulk region $0 < y < L$.  
The KK expansion of $\tilde \Psi $ in the type 1A is given by
\begin{align}
\begin{pmatrix} \tilde u_{R} (x,y) \cr \noalign{\kern 5pt} \tilde d_{R} (x,y) \end{pmatrix} 
&= \sum_{n=-\infty}^{\infty } \psi_{R}^{(n)}  (x) 
\frac{1}{\sqrt{\pi R}}
\begin{pmatrix} \cos \Big[ \myfrac{ny}{R} - \onehalf \theta (y) \Big] \cr 
 \sin \Big[ \myfrac{ny}{R} - \onehalf \theta (y) \Big] \end{pmatrix}, \cr
\noalign{\kern 5pt}
\begin{pmatrix} \tilde u_{L} (x,y) \cr \noalign{\kern 5pt} \tilde d_{L} (x,y) \end{pmatrix} 
&= \sum_{n=-\infty}^{\infty } \psi_{L}^{(n)}  (x) 
\frac{1}{\sqrt{\pi R}}
\begin{pmatrix} - \sin \Big[ \myfrac{ny}{R} - \onehalf \theta (y) \Big] \cr 
 \cos \Big[ \myfrac{ny}{R} - \onehalf \theta (y) \Big] \end{pmatrix}.
\label{fermionKKflat1}
\end{align}
In the original gauge it becomes 
\begin{align}
\underline{\hbox{type 1A}}:  ~~\begin{pmatrix} u_{R} (x,y) \cr \noalign{\kern 5pt} d_{R} (x,y) \end{pmatrix} 
&= \sum_{n=-\infty}^{\infty } \psi_{R}^{(n)}  (x) 
\frac{1}{\sqrt{\pi R}}
\begin{pmatrix} \cos  \myfrac{ny}{R}  \cr 
 \sin \myfrac{ny}{R}  \end{pmatrix}, \cr
\noalign{\kern 5pt}
\begin{pmatrix}u_{L} (x,y) \cr \noalign{\kern 5pt} d_{L} (x,y) \end{pmatrix} 
&= \sum_{n=-\infty}^{\infty } \psi_{L}^{(n)}  (x) 
\frac{1}{\sqrt{\pi R}}
\begin{pmatrix} - \sin \myfrac{ny}{R} \cr 
 \cos  \myfrac{ny}{R}  \end{pmatrix}.
\label{fermionKKflat2}
\end{align}
$\psi_{R}^{(n)}  $ and $\psi_{L}^{(n)} $ combine to form
the $\psi^{(n)}  (x) $ mode, whose mass is given by 
$m_{n}(\theta_{H}) = R^{-1} \big| n + \frac{\theta_{H}}{2 \pi} \big|$. 
The spectrum is periodic in $\theta_H$ with period $2\pi$.
The KK expansion for type 1B is obtained by interchanging left-handed and right-handed components 
in (\ref{fermionKKflat2}).

For $\Psi$ in type 2A the KK expansion is
\begin{align}
\underline{\hbox{type 2A}}:  ~~\begin{pmatrix} u_{R} (x,y) \cr \noalign{\kern 5pt} d_{R} (x,y) \end{pmatrix} 
&= \sum_{n=-\infty}^{\infty } \psi_{R}^{( n + \frac{1}{2} )} (x) 
\frac{1}{\sqrt{\pi R}}
\begin{pmatrix} \cos  \myfrac{(n + \onehalf) y}{R}  \cr 
 \sin \myfrac{(n + \onehalf) y}{R}  \end{pmatrix}, \cr
\noalign{\kern 5pt}
\begin{pmatrix}u_{L} (x,y) \cr \noalign{\kern 5pt} d_{L} (x,y) \end{pmatrix} 
&= \sum_{n=-\infty}^{\infty } \psi_{L}^{(n + \frac{1}{2} )} (x) 
\frac{1}{\sqrt{\pi R}}
\begin{pmatrix} -\sin \myfrac{(n + \onehalf) y}{R} \cr 
\cos  \myfrac{(n + \onehalf) y}{R}  \end{pmatrix}.
\label{fermionKKflat3}
\end{align}
$\ \psi_{R}^{( n + \frac{1}{2})}$ and 
$\psi_{L}^{( n + \frac{1}{2} )}$  combine to form
the $\psi^{( n + \frac{1}{2} ) }(x)$ mode, whose mass is given by 
$m_{n + \frac{1}{2}}(\theta_{H}) = R^{-1} \big| n + \onehalf + \frac{\theta_{H}}{2 \pi} \big|$. 
The KK expansion for type 2B is obtained by interchanging left-handed and right-handed components 
in (\ref{fermionKKflat3}).


Next we examine $SU(2)$ GHU in the RS space whose metric is given by \cite{RS1999}
\begin{align}
ds^2= e^{-2\sigma(y)} \eta_{\mu\nu}dx^\mu dx^\nu+dy^2
\label{RSmetric1}
\end{align}
where $\eta_{\mu\nu}=\mbox{diag}(-1,+1,+1,+1)$, $\sigma(y)=\sigma(y+ 2L)=\sigma(-y)$ 
and $\sigma(y)=ky$ for $0 \le y \le L$.  It has the same topology as $M^4 \times (S^{1}/Z_{2})$.
In the fundamental region $0 \le y \le L$ the metric can be written, in terms of the conformal coordinate 
$z = e^{ky}$, as
\begin{align}
ds^2=  \frac{1}{z^2} \bigg(\eta_{\mu\nu}dx^{\mu} dx^{\nu} + \frac{dz^2}{k^2}\bigg) 
\quad ( 1 \le z \le z_L= e^{kL}) ~.
\label{RSmetric2}
\end{align}
$z_L$ is called the warp factor of the RS space.
The action in RS is 
\begin{align}
I_\RS &= \int d^5 x  \sqrt{- \det G}   \, {\cal L}_\RS ~, \cr
\noalign{\kern 5pt}
{\cal L}_\RS & = - \frac{1}{2} \Tr  F_{MN} F^{MN} 
+  \Psibar {\cal D} (c) \Psi   ~, \cr
\noalign{\kern 5pt}
{\cal D} (c) &=  \gamma^A {e_A}^M
\bigg( D_M+\frac{1}{8}\omega_{MBC}[\gamma^B,\gamma^C]  \bigg) - c \, \sigma' 
\label{RSaction}
\end{align}
where $\sigma'  (y)= k$ for $0 \le y \le L$.
Note ${\cal L}_\RS (x^\mu, y) = {\cal L}_\RS  (x^\mu, -y) = {\cal L}_\RS (x^\mu, y + 2L)$.
Fields $A_{M}$ and $\Psi$ satisfy the same boundary conditions (\ref{BC1}) as in the flat spacetime.
The dimensionless bulk mass parameter $c$ in ${\cal D} (c)$ controls the mass and wave function of the fermion field.
The KK mass scale is given by
\begin{align}
m_{\KK} &= \frac{\pi k}{z_{L}-1} 
\label{KKscale1}
\end{align}
which becomes $1/R$ in the flat spacetime limit $k \go 0$.

In the KK expansion in the $z$ coordinate, $A_{z}^{a} (x,z) = k^{-1/2} \sum A_{z}^{a (n)} (x)  h_{n}(z)$, 
the zero mode  $A_{z}^{2 (0)}$ has a wave function $h_{0} (z) = \sqrt{2/(z_{L}^{2}-1)} \,  z$.
In the $y$-coordinate   $A_{y}^{2 (0)}$ has a wave function
$v_{0} (y) = k e^{ky} h_{0}(z) $ for $0 \le y \le L$ and $v_{0} (-y) = v_{0} (y) =  v_{0} (y+2L)$.
The AB phase $\theta_{H}$ in (\ref{ABphase1}) becomes
\begin{align}
\theta_{H} &= \frac{\la A_{z}^{2 (0)} \ra}{f_{H}} ~, ~~
f_{H} = \frac{1}{g_{4}} \sqrt{ \frac{2k}{L(z_{L}^{2} - 1)} } ~.
\label{ABphase3}
\end{align}
The twisted gauge \cite{Falkowski2007, HS2007}, in which $\tilde \theta_{H} = 0$, is related to the original 
gauge by a large gauge transformation
\begin{align}
\Omega (z) & = e^{i \theta (z) \tau^{2}/2} ~, ~~
\theta (z) =  \theta_{H} \, \frac{z_{L}^{2} - z^{2}}{z_{L}^{2} - 1} ~.
\label{largeGT1}
\end{align}
In the $y$-coordinate it becomes
\begin{align}
\Omega (y) & = \exp \bigg\{ i \theta_{H} \sqrt{\frac{2}{z_{L}^{2} -1}} \int_{y}^{L} dy \, v_{0} (y) \cdot \frac{\tau^{2}}{2} \bigg\}.
\label{largeGT2}
\end{align}

In the twisted gauge $\tilde A_{\mu}^{1,3} (x,z)$ satisfy free equations in $1 \le z \le z_L$ and boundary conditions (\ref{BC2}).
The mass spectrum $\{ m_n (\theta_H)  = k \lambda_n  (\theta_H) \}$ ($ \lambda_0 <  \lambda_1 <  \lambda_2 < \cdots$) 
is given by
\begin{align}
Z_\mu^{(n)} : ~ S C' (1; \lambda_{n}) + \lambda_{n} \sin^{2} \theta_{H} = 0 
\label{RSgaugespectrum1}
\end{align}
where $S (z; \lambda)$ and  $C (z; \lambda)$ are expressed in terms of Bessel functions and 
are given by (\ref{functionA1}).
The KK expansions in the twisted gauge in the region $1 \le z \le z_L$ are written 
as\footnote{Note a change in the normalization of mode functions.  $\tilde{\bf h}_n (z)$ in the present paper corresponds to
$\sqrt{kL} \, \tilde{\bf h}_n (z)$ in Ref.\ \cite{anomalyFlow2022}.}
\begin{align}
\begin{pmatrix} \tilde A_{\mu}^{1} (x,z) \cr \noalign{\kern 5pt} \tilde A_{\mu}^{3} (x,z) \end{pmatrix} 
&= \frac{1}{\sqrt{L}}  \sum_{n=0}^{\infty } Z_{\mu}^{(n)} (x) \, \tilde{\bf h}_n (z) ~,~~
 \tilde{\bf h}_n (z) = \begin{pmatrix} \tilde h_n(z) \cr \tilde k_n (z) \end{pmatrix}
\label{RSgaugeKK1}
\end{align}
where the mode functions $\tilde{\bf h}_n (z)$ are given in (\ref{RSgaugeKKB1}).
In the original gauge the KK expansions of $A_\mu^{1,3} (x,y)$ become
\begin{align}
\begin{pmatrix} A_{\mu}^{1} (x,y) \cr \noalign{\kern 5pt}  A_{\mu}^{3} (x,y) \end{pmatrix} 
&=  \frac{1}{\sqrt{L}} \sum_{n=0}^{\infty } Z_{\mu}^{(n)} (x) \,  \begin{pmatrix}  h_n(y) \cr  k_n (y) \end{pmatrix}, \cr
\noalign{\kern 5pt}
\begin{pmatrix}  h_n(y) \cr  k_n (y) \end{pmatrix} 
&= \begin{pmatrix}  - h_n(- y) \cr  k_n (- y) \end{pmatrix}
= \begin{pmatrix}  h_n(y+ 2L) \cr  k_n (y + 2L) \end{pmatrix} \cr 
\noalign{\kern 5pt}
&= \begin{pmatrix} \cos \theta (z) & \sin \theta (z) \cr - \sin \theta (z) & \cos \theta (z) \end{pmatrix} 
\begin{pmatrix} \tilde h_n(z) \cr \tilde k_n (z) \end{pmatrix} \quad {\rm for~} 0 \le y \le L ~.
\label{RSgaugeKK2}
\end{align}

For a fermion field $\Psi(x,z)$ it is most convenient to express its
KK expansion for $\check \Psi(x,z) = z^{-2} \Psi(x,z)$.
Equations of motion in the region $1 \le z \le z_L$  become
\begin{align}
& -k D_- (c) \, \check \Psi_R + \sigma^\mu \dd_\mu \check \Psi_L = 0 ~,~
 -k D_+ (c) \, \check \Psi_L + \bar \sigma^\mu \dd_\mu \check \Psi_R = 0 ~,    \cr
\noalign{\kern 3pt}
&\quad \sigma^\mu = (I_2, \vec \sigma) ~,~ \bar \sigma^\mu = (-I_2, \vec \sigma) ~,~
D_\pm (c) = \pm \frac{\dd}{\dd z} + \frac{c}{z} ~.
\label{RSfermionEq1}
\end{align}
In the presence of gauge fields $\dd_M$ is replaced by $\dd_M - i g_A A_M$.
The Neumann boundary conditions at $z= (z_{0}, z_{1}) = (1, z_{L})$,  
corresponding to even parity,  for left- and right-handed components are
given by $D_{+} (c) \check \Psi_{L} \big|_{z_{j}} =0$ and $ D_{-}(c)  \check \Psi_{R} \big|_{z_{j}} =0$.

\begin{figure}[tbh]
\centering
\includegraphics[width=100mm]{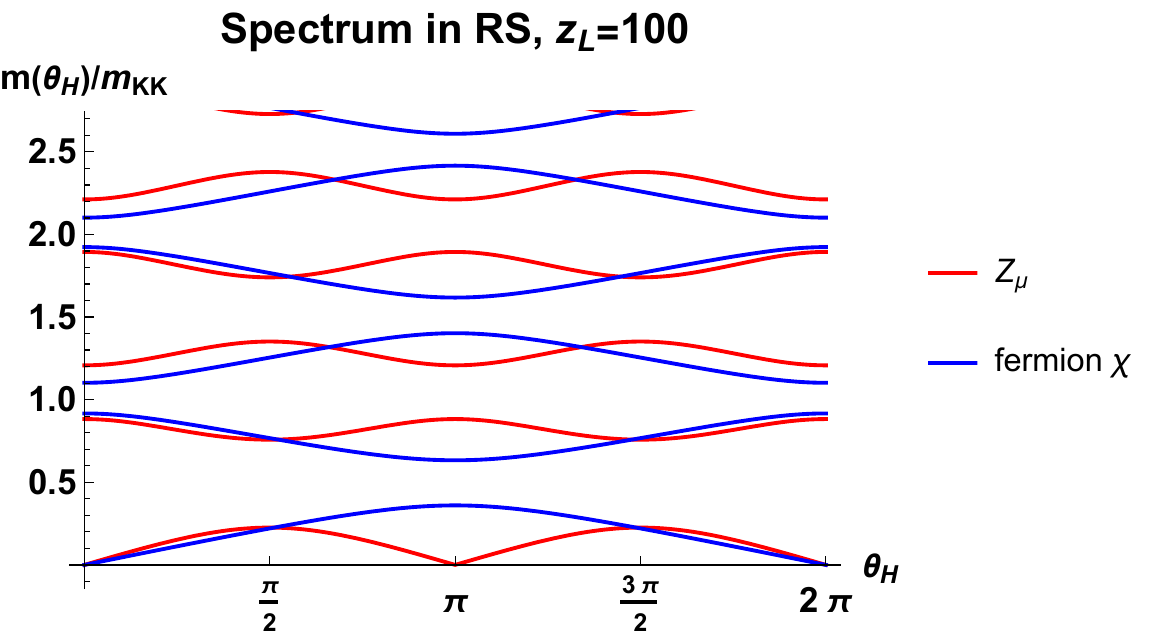}
\caption{The mass spectrum of gauge fields $Z_{\mu}^{(n)}$ and fermion  fields $\chi^{(n)}$ (type 1A)
in the RS warped space is displayed.  The warp factor is $z_L = 100$ and the bulk mass parameter 
of $\Psi$ is $c=0.25$.  There is no level crossing in the spectrum.
}   
\label{fig:RSspectrum}
\end{figure}

The spectrum of the KK modes of the fermion field $\Psi$  is determined by
\begin{align}
\chi^{(n)}  : ~
\begin{cases}
S_{L} S_{R}(1; \lambda_{n}, c) +  \sin^{2} \onehalf \theta_{H} = 0  & \hbox{for type 1A/B} \cr
\noalign{\kern 5pt}
S_{L} S_{R}(1; \lambda_{n}, c) +  \cos^{2} \onehalf \theta_{H} = 0  & \hbox{for type 2A/B} 
\end{cases}
\label{RSfermispectrum1}
\end{align}
where functions $S_{L/R} (z;  \lambda, c)$ are given in (\ref{functionA3}).
The spectrum is periodic in $\theta_H$ with period $2 \pi$.  
A massless mode appears at $\theta_H=0$ for type 1A and 1B,  whereas it appears
at $\theta_H = \pi$ for type 2A and 2B.  There is no level crossing in the spectrum except for the case $c=0$.
The spectra of the gauge fields (\ref{RSgaugespectrum1}) and fermion fields (\ref{RSfermispectrum1})
are displayed in Figure \ref{fig:RSspectrum}.

The KK expansion of the fermion field $\Psi$ in the twisted gauge in the region $1 \le z \le z_L$ is expressed as
\begin{align}
\begin{pmatrix} \tilde{\check u}_{R} (x,z) \cr \noalign{\kern 5pt} \tilde{\check d}_{R} (x,z) \end{pmatrix} 
&= \sqrt{k} \sum_{n= 0}^\infty   \chi_{R}^{(n)} (x) \, \tilde{\bf f}_{Rn} (z) ~,~
\tilde{\bf f}_{Rn} (z)  = \begin{pmatrix} \tilde f_{Rn} (z) \cr \tilde g_{Rn} (z) \end{pmatrix},  \cr
\noalign{\kern 5pt}
\begin{pmatrix} \tilde{\check u}_{L} (x,z) \cr \noalign{\kern 5pt} \tilde{\check d}_{L} (x,z) \end{pmatrix} 
&= \sqrt{k} \sum_{n= 0}^\infty   \chi_{L}^{(n)} (x) \, \tilde{\bf f}_{Ln} (z) ~, ~
\tilde{\bf f}_{Ln} (z)  = \begin{pmatrix} \tilde f_{Ln} (z) \cr \tilde g_{Ln} (z) \end{pmatrix}.
\label{RSfermionKK1}
\end{align}
The  mode functions $\tilde{\bf f}_{Rn} (z)$ and $\tilde{\bf f}_{Ln} (z)$ for type 1A are given in (\ref{RSfermionKKB1}).
In the original gauge the expansions of $\check u (x,y)$ and $\check d (x,y)$ become
\begin{align}
\begin{pmatrix}{\check u}_{R} (x,y) \cr \noalign{\kern 2pt} {\check d}_{R} (x,y) \end{pmatrix} 
&= \sqrt{k} \sum_{n= 0}^\infty   \chi_{R}^{(n)} (x)    \begin{pmatrix}  f_{Rn} (y) \cr   g_{Rn} (y) \end{pmatrix},  \cr
\noalign{\kern 5pt}
\begin{pmatrix}  {\check u}_{L} (x,y) \cr \noalign{\kern 2pt} {\check d}_{L} (x,y) \end{pmatrix} 
&= \sqrt{k} \sum_{n= 0}^\infty   \chi_{L}^{(n)} (x)  \begin{pmatrix}   f_{Ln} (y) \cr   g_{Ln} (y) \end{pmatrix},
\label{RSfermionKK2}
\end{align}
where
\begin{align}
\underline{\hbox{type 1A}} ~~&  \cr
\noalign{\kern 5pt}
\begin{pmatrix}  f_{Rn} (y) \cr   g_{Rn} (y) \end{pmatrix}
&=  \begin{pmatrix}  f_{Rn} (- y) \cr   - g_{Rn} (- y) \end{pmatrix} 
=  \begin{pmatrix}  f_{Rn} (y+ 2L) \cr   g_{Rn} (y+ 2L) \end{pmatrix}  \cr
\noalign{\kern 5pt}
&=  \begin{pmatrix} \cos \onehalf \theta (z) & - \sin \onehalf \theta (z) \cr 
\noalign{\kern 2pt}
 \sin\onehalf  \theta (z) & \cos \onehalf \theta (z) \end{pmatrix} 
 \begin{pmatrix}  \tilde f_{Rn} (z) \cr   \tilde g_{Rn} (z) \end{pmatrix}  \quad {\rm for~} 0 \le y \le L ~, \cr
\noalign{\kern 8pt}
\begin{pmatrix}  f_{Ln} (y) \cr   g_{Ln} (y) \end{pmatrix}
&=  \begin{pmatrix}  - f_{Ln} (- y) \cr    g_{Ln} (- y) \end{pmatrix} 
=  \begin{pmatrix}  f_{Ln} (y+ 2L) \cr   g_{Ln} (y+ 2L) \end{pmatrix}  \cr
\noalign{\kern 5pt}
& =  \begin{pmatrix} \cos \onehalf \theta (z) & - \sin \onehalf \theta (z) \cr 
\noalign{\kern 2pt}
 \sin\onehalf  \theta (z) & \cos \onehalf \theta (z) \end{pmatrix} 
 \begin{pmatrix}  \tilde f_{Ln} (z) \cr   \tilde g_{Ln} (z) \end{pmatrix}  \quad {\rm for~} 0 \le y \le L ~, \cr
 \noalign{\kern 8pt}
\underline{\hbox{type 2A}} ~~&  \cr
\noalign{\kern 5pt}
\begin{pmatrix}  f_{Rn} (y) \cr   g_{Rn} (y) \end{pmatrix}
&=  \begin{pmatrix}  f_{Rn} (- y) \cr   - g_{Rn} (- y) \end{pmatrix} 
=  \begin{pmatrix}  - f_{Rn} (y+ 2L) \cr   - g_{Rn} (y+ 2L) \end{pmatrix} ,   \cr
\noalign{\kern 5pt}
\begin{pmatrix}  f_{Ln} (y) \cr   g_{Ln} (y) \end{pmatrix}
&=  \begin{pmatrix}  - f_{Ln} (- y) \cr    g_{Ln} (- y) \end{pmatrix} 
=  \begin{pmatrix}  - f_{Ln} (y+ 2L) \cr   - g_{Ln} (y+ 2L) \end{pmatrix} .
\label{RSfermionKK3}
\end{align}
For type 1B (2B), the parity of $f_{R/Ln}, g_{R/Ln}$ is reversed compared to type 1A (2A).

\section{Anomalies}

Doublet fermions in type 1A or 1B are chiral at $\theta_H =0$.  Massless modes appear for 
right-handed $u$ and left-handed $d$ (left-handed $u$ and right-handed $d$) for type 1A (1B).
They become massive as $\theta_H$ varies, and their gauge couplings become purely vector-like 
at $\theta_H=\pi$.  Chiral anomalies exist at $\theta_H=0$, smoothly vary as $\theta_H$ in the RS space, and 
vanish at $\theta_H=\pi$.  This phenomenon is called the anomaly flow by an AB phase \cite{anomalyFlow2022}.

Chiral anomalies arise from triangular loop diagrams.  Gauge couplings of fermions have been
obtained in Ref.\cite{anomalyFlow2022}.   Substituting the KK expansions (\ref{RSgaugeKK2}) and
(\ref{RSfermionKK2}) into 
\begin{align}
g_A  \int_1^{z_L} \frac{dz}{k} \, \Big\{ 
{\check \Psi}_R^\dagger \bar \sigma^\mu  A_\mu {\check \Psi}_R
-{\check \Psi}_L^\dagger \sigma^\mu  A_\mu {\check \Psi}_L \Big\} ~,
\label{RSgagueInt1}
\end{align}
one finds that the couplings in 
\begin{align}
&\frac{g_4}{2} \sum_{n=0}^\infty \sum_{\ell = 0}^\infty \sum_{m=0}^\infty Z_\mu^{(n)} (x)
\Big\{ t^R_{n\ell m} \, \chi_R^{(\ell)} (x)^\dagger \bar \sigma^\mu \chi_R^{(m)} (x) 
+ t^L_{n\ell m} \, \chi_L^{(\ell)} (x)^\dagger  \sigma^\mu \chi_L^{(m)} (x) \Big\} 
\label{RSgaugeCoupling1}
\end{align}
are given by
\begin{align}
t^R_{n\ell m} &=  \int_1^{z_L} dz \, \Big\{ 
\tilde h_n(z) \big( \tilde f_{R\ell}^* (z) \tilde g_{R m} (z)  + \tilde g_{R\ell}^* (z) \tilde f_{R m} (z) \big) \cr
\noalign{\kern 3pt}
&\hskip 3.cm
+ \tilde k_n(z) \big( \tilde f_{R\ell}^* (z) \tilde f_{R m} (z)  - \tilde g_{R\ell}^* (z) \tilde g_{R m} (z) \big) \Big\}  \cr
\noalign{\kern 5pt}
&= \frac{k }{2} \int_{-a}^{2L - a} dy \,  e^{\sigma(y)} \Big\{ 
h_n(y) \big( f_{R\ell}^* (y) g_{R m} (y)  + g_{R\ell}^* (y) f_{R m} (y) \big) \cr
\noalign{\kern 3pt}
&\hskip 3.cm
+ k_n(y) \big( f_{R\ell}^* (y) f_{R m} (y)  - g_{R\ell}^* (y) g_{R m} (y) \big) \Big\} , \cr
\noalign{\kern 5pt}
t^L_{n\ell m} &=  -  \int_1^{z_L} dz \, \Big\{ 
h_n(z) \big( f_{L\ell}^* (z) g_{L m} (z)  + g_{L\ell}^* (z) f_{L m} (z) \big) \cr
\noalign{\kern 3pt}
&\hskip 3.cm
+ k_n(z) \big( f_{L\ell}^* (z) f_{L m} (z)  - g_{L\ell}^* (z) g_{L m} (z) \big) \Big\}  \cr
\noalign{\kern 5pt}
&=-  \frac{k}{2} \int_{-a}^{2L - a} dy \,  e^{\sigma(y)} \Big\{ 
h_n(y) \big( f_{L\ell}^* (y) g_{L m} (y)  + g_{L\ell}^* (y) f_{L m} (y) \big) \cr
\noalign{\kern 3pt}
&\hskip 3.cm
+ k_n(y) \big( f_{L\ell}^* (y) f_{L m} (y)  - g_{L\ell}^* (y) g_{L m} (y) \big) \Big\} .
\label{RSgaugeCoupling2}
\end{align}
The couplings $t^R_{n\ell m}$ and $t^L_{n\ell m}$ are gauge-invariant.  
In the integral formulas in the $y$-coordinate the constant $a$ is arbitrary as the integrands
are periodic functions with period $2L$.
It is convenient to take $0 < a < L$ in the following discussions.
We note that the couplings $t^{R/L}_{n\ell m}$ depend not only on $\theta_H$ and  $z_L$, but also
on the bulk mass parameter $c$ of the fermion field $\Psi$.

The anomaly coefficient associated with the three legs  of 
$Z_{\mu_1}^{(n_1)} Z_{\mu_2}^{(n_2)} Z_{\mu_3}^{(n_3)}$ is given by 
\begin{align}
&a_{n_{1} n_{2} n_{3}} = a_{n_{1} n_{2} n_{3}}^{R} + a_{n_{1} n_{2} n_{3}}^{L} ~, \cr
\noalign{\kern 5pt}
&a_{n_{1} n_{2} n_{3}}^{R} = \Tr T^{R}_{n_{1}} T^{R}_{n_{2}} T^{R}_{n_{3}} ~,~~
(T^{R}_{n} )_{m \ell} = t^{R}_{n m \ell} ~, \cr
\noalign{\kern 5pt}
&a_{n_{1} n_{2} n_{3}}^{L} = \Tr T^{L}_{n_{1}} T^{L}_{n_{2}} T^{L}_{n_{3}} ~,~~
(T^{L}_{n} )_{m \ell} = t^{L}_{n m \ell} ~.
\label{RSAnomaly1}
\end{align}
The anomaly coefficient $a_{n_{1} n_{2} n_{3}}$ depends on  $\theta_{H}$, exhibiting the anomaly flow.
It has been observed by numerical evaluation in Ref.\cite{anomalyFlow2022} that 
$a_{n_{1} n_{2} n_{3}}$ does not depend on the bulk mass parameter $c$,
though $a_{n_{1} n_{2} n_{3}}^{R}$ and $a_{n_{1} n_{2} n_{3}}^{L}$ do depend on $c$.
We are going to show that $a_{n_{1} n_{2} n_{3}} (\theta_H, z_L)$ is expressed
in terms of the values of the wave functions $k_{n_j} (y)$ at $y=0$ and $y=L$.

To see it we insert the formulas for $t^{R/L}_{n\ell m}$ in (\ref{RSgaugeCoupling2}) into (\ref{RSAnomaly1}),
and rearrange the traces.
\begin{align}
&a_{n_{1} n_{2} n_{3}} = \Big( \frac{k }{2} \Big)^{3} \int \int \int_{-a}^{2L-a} dy_1 dy_2 dy_3 \,
e^{\sigma (y_1) + \sigma (y_2) + \sigma (y_3)} \cr
\noalign{\kern 5pt}
&\quad
\times \Big[ ~  k_1 k_2 k_3 \big\{ A_R(1,2) A_R(2,3) A_R (3,1) - B_R(1,2) B_R(2,3) B_R (3,1)  \cr
\noalign{\kern 2pt}
&\hskip 2.2cm
+ B_L(1,2) B_L(2,3) B_L (3,1) - A_L(1,2) A_L(2,3) A_L (3,1) \big\} \cr
\noalign{\kern 2pt}
&\quad \quad
+ k_1 h_2 h_3 \big\{ A_R(1,2) B_R(2,3) A_R (3,1) - B_R(1,2) A_R(2,3) B_R (3,1)  \cr
\noalign{\kern 2pt}
&\hskip 2.2cm
+ B_L(1,2) A_L(2,3) B_L (3,1) - A_L(1,2) B_L(2,3) A_L (3,1) \big\} \cr
\noalign{\kern 2pt}
&\quad \quad
+ h_1 k_2 h_3  \big\{ A_R(1,2) A_R(2,3) B_R (3,1) - B_R(1,2) B_R(2,3) A_R (3,1)  \cr
\noalign{\kern 2pt}
&\hskip 2.2cm
+ B_L(1,2) B_L(2,3) A_L (3,1) - A_L(1,2) A_L(2,3) B_L (3,1) \big\} \cr
\noalign{\kern 2pt}
&\quad \quad
+ h_1 h_2 k_3  \big\{ B_R(1,2) A_R(2,3) A_R (3,1) - A_R(1,2) B_R(2,3) B_R (3,1)  \cr
\noalign{\kern 2pt}
&\hskip 2.2cm
+ A_L(1,2) B_L(2,3) B_L (3,1) - B_L(1,2) A_L(2,3) A_L (3,1) \big\}  ~ \Big]
\label{RSAnomaly2}
\end{align}
where
\begin{align}
&k_j = k_{n_j} (y_j) ~, ~~ h_j = h_{n_j} (y_j) ~, \cr
\noalign{\kern 5pt}
&\begin{pmatrix} A_{R/L} (j, k) \cr B_{R/L} (j, k) \end{pmatrix}
 = \begin{pmatrix} A_{R/L} \cr B_{R/L}  \end{pmatrix}  (y_j, y_k) 
 = \sum_{n=0}^\infty  \begin{pmatrix}  f_{R/L n} (y_j) f_{R/L n}^* (y_k)  \cr
g_{R/L n} (y_j) g_{R/L n}^* (y_k)  \end{pmatrix} .
\label{RSdef1}
\end{align}
Eqs.\ (\ref{RSfermionKK1}) and (\ref{RSfermionKK2}) and the orthonormality relations of the mode functions 
imply that 
\begin{align}
&\begin{pmatrix} \check u_{R/L} (x, y) \cr \check d_{R/L} (x, y) \end{pmatrix} 
= \frac{k}{2} \int_{-a}^{2L -a} dy' \, e^{\sigma(y')} 
\begin{pmatrix} A_{R/L} &C_{R/L} \cr D_{R/L} & B_{R/L} \end{pmatrix} (y, y') 
\begin{pmatrix} \check u_{R/L} (x, y') \cr \check d_{R/L} (x, y') \end{pmatrix} , \cr
\noalign{\kern 5pt}
&\hskip 2.cm  
\begin{pmatrix} C_{R/L} \cr D_{R/L}  \end{pmatrix}  (y, y') 
 = \sum_{n=0}^\infty  \begin{pmatrix}  f_{R/L n} (y) g_{R/L n}^* (y')  \cr
g_{R/L n} (y) f_{R/L n}^* (y')  \end{pmatrix} .
\label{RSdef2}
\end{align}
We have made use of the relation $C_{R/L} = D_{R/L} =0$ in deriving (\ref{RSAnomaly2}).
With the choice of the AB phase $\theta_H$ in (\ref{ABphase3}) all mode functions $\{ f_{Rn} (y) \}$ etc.\ can be
taken to be real so that $A_{R/L} (y, y') = A_{R/L} (y', y) $ and $B_{R/L} (y, y') = B_{R/L} (y', y) $.

In addition to the relation (\ref{RSdef2}), $A_{R/L}$ and $B_{R/L}$ must satisfy the parity relations and
boundary conditions of the mode functions.  With $(y_0, y_1) = (0, L)$
\begin{align}
&
\underline{\hbox{type 1A}} :  \cr
\noalign{\kern 5pt}
&\begin{pmatrix} A_R \cr B_R \cr A_L \cr B_L \end{pmatrix} (y_j -y, y') 
= \begin{pmatrix} A_R \cr -B_R \cr - A_L \cr B_L \end{pmatrix}  (y_j +y, y') ~, \cr
\noalign{\kern 5pt}
&\begin{pmatrix} \hat D_- (c) A_R (y, y')  \cr \hat D_+ (c) B_L (y, y') \end{pmatrix}_{y=\ep, L -\ep} = 0 ~ , ~
\hat D_\pm (c) = \pm \frac{\dd}{\dd y} + c \, k ~, \cr
\noalign{\kern 5pt}
&B_R (y_j , y') = A_L  (y_j , y') = 0 ~, \cr
\noalign{\kern 5pt}
&\underline{\hbox{type 2A}} :  \cr
\noalign{\kern 5pt}
&\begin{pmatrix} A_R \cr B_R \cr A_L \cr B_L \end{pmatrix} (y_j -y, y') 
=
\begin{pmatrix} (-1)^j A_R \cr (-1)^{j+1}  B_R \cr (-1)^{j+1} A_L \cr (-1)^j B_L \end{pmatrix}  (y_j +y, y') ~, \cr
\noalign{\kern 5pt}
&\begin{pmatrix} \hat D_- (c) A_R (y, y')  \cr \hat D_+ (c) B_L (y, y') \end{pmatrix}_{y=\ep} = 
\begin{pmatrix} \hat D_- (c) B_R (y, y')  \cr \hat D_+ (c) A_L (y, y') \end{pmatrix}_{y= L -\ep} =0 ~ , \cr
\noalign{\kern 5pt}
&B_R (0 , y') = A_R (L , y') = A_L  (0 , y') = B_L  (L , y') =0 ~.
\label{RSBC2}
\end{align}
The condition for type 1B (2B) are obtained by interchanging $R$ (right-handed) and $L$ (left-handed)  in those 
for type 1A (2A).  For $c \not= 0$,  parity even components of $A_{R/L}$ and $B_{R/L}$ functions exhibit 
the cusp behavior at $y, y' = 0, \pm L, \cdots$.

It is not easy to explicitly write down $A_{R/L}(y,y')$ and $B_{R/L} (y,y')$ functions for $c\not= 0$ which satisfy
the relations in both (\ref{RSdef2}) and (\ref{RSBC2}).
In the previous paper \cite{anomalyFlow2022} it has been recognized that the anomaly coefficient 
$a_{n_1 n_2 n_3}$ in (\ref{RSAnomaly2}) is independent of $c$.  With this observation we shall
derive an analytical expression for $a_{n_1 n_2 n_3}$ by evaluating it in the case $c=0$.
We will confirm later that numerically evaluated $a_{n_1 n_2 n_3}$ for  $c \not= 0$ agrees with
the analytical formula.

Fermion wave functions for $c=0$ are expressed in terms of trigonometric functions.
They are summarized in Appendix B.3.
Inserting the wave functions in (\ref{RSfermionKKB3}) into $A_R(z,z') = \sum f_{Rn} (z) f_{Rn}^* (z')$,
for instance, one finds for type 1A that, for $1 \le z, z' \le z_L$, 
\begin{align}
&A_R(z,z')^{c=0}  \cr
&= \frac{1}{z_L -1} \sum_{n=-\infty}^\infty 
\cos \Big( n \pi \frac{z-z_L}{z_L-1} + \alpha(z) \Big) \cos \Big( n \pi \frac{z'-z_L}{z_L-1} + \alpha(z') \Big) \cr
\noalign{\kern 5pt}
&= \delta_{2(z_L-1)} (z-z' ) \cos\big\{ \alpha(z) - \alpha(z') \big\} 
+ \delta_{2(z_L-1)} (z+z'-2) \cos\big\{ \alpha(z) + \alpha(z') \big\}  \cr
\noalign{\kern 5pt}
&=  \delta_{2(z_L-1)} (z-z' ) + \delta_{2(z_L-1)} (z+z'-2) ~, \cr
\noalign{\kern 5pt}
&\quad 
\alpha(z) = \frac{1}{2} \Big\{ \theta_H \,  \frac{z-z_L}{z_L-1} + \theta (z) \Big\} ~, ~ \alpha(1) = \alpha(z_L) =0 ~.
\label{RSdef3}
\end{align}
Here $\delta_L (x) = \sum_n \delta(x - nL)$.
With the extension (\ref{RSfermionKK3}) in the $y$-coordinate and similar manipulation one finds that
\begin{align}
&\underline{\hbox{type 1A}, ~ c=0} \cr
\noalign{\kern 5pt}
&A_R(y,y') = B_L (y,y')
= \frac{e^{-\sigma (y)}}{k} \big\{ \delta_{2L} (y -y') + \delta_{2L} (y +y')  \big\} ~, \cr
\noalign{\kern 5pt}
&B_R(y,y') = A_L (y,y') 
= \frac{e^{-\sigma (y)}}{k} \big\{ \delta_{2L} (y -y')  - \delta_{2L} (y +y') \big\} ~,
\label{completeness1}
\end{align}
Formulas for type 1B are obtained by interchanging $R$ and $L$.

For fermions in type 2A, one finds for $1 \le z, z' \le z_L$ that
\begin{align}
&A_R(z,z')^{c=0}  \cr
&= \frac{1}{z_L -1} \sum_{n=-\infty}^\infty 
\sin \Big( n \pi \frac{z-z_L}{z_L-1} + \beta(z) \Big) \sin \Big( n \pi \frac{z'-z_L}{z_L-1} + \beta(z') \Big) \cr
\noalign{\kern 5pt}
&= \delta_{2(z_L-1)} (z-z' ) \cos\big\{ \beta(z) - \beta(z') \big\} 
- \delta_{2(z_L-1)} (z+z'-2) \cos\big\{ \beta(z) + \beta(z') \big\} ~,  \cr
\noalign{\kern 5pt}
&\quad 
\beta(z) = \frac{1}{2} \Big\{ (\theta_H + \pi)  \,  \frac{z-z_L}{z_L-1} + \theta (z) \Big\} ~, ~ 
\beta(1) =  - \frac{1}{2} \pi ~, ~\beta(z_L) =0 ~.
\label{RSdef4}
\end{align}
Noting the relations in (\ref{RSfermionKK3}), one finds  in the $y$-coordinate that 
\begin{align}
&\underline{\hbox{type 2A}, ~ c=0} \cr
\noalign{\kern 5pt}
&A_R(y,y') = B_L (y,y')
= \frac{e^{-\sigma (y)}}{k} \big\{ \hat \delta_{2L} (y -y') + \hat \delta_{2L} (y +y')  \big\} ~, \cr
\noalign{\kern 5pt}
&B_R(y,y') = A_L (y,y') 
= \frac{e^{-\sigma (y)}}{k} \big\{ \hat \delta_{2L} (y -y')  - \hat \delta_{2L} (y +y') \big\} ~, \cr
\noalign{\kern 8pt}
&\hat \delta_{2L} (y) =   \delta_{4L} (y) - \delta_{4L} (y - 2L) ~.
\label{completeness2}
\end{align}
Formulas for type 2B are obtained by interchanging $R$ and $L$.

We insert the expressions (\ref{completeness1}) or  (\ref{completeness2}) into (\ref{RSAnomaly2}).
There appear products of three delta functions in the integrand.
Take $0 < a < L$.  Then in the integration range $-a \le y_1, y_2, y_3 \le 2L - a$, products of delta functions
reduce to
\begin{align}
&\begin{matrix} \delta_{2L} (y_1 - y_2)  \delta_{2L} (y_2 - y_3)  \delta_{2L} (y_3 + y_1) \cr
\noalign{\kern 5pt}
\delta_{2L} (y_1 + y_2)  \delta_{2L} (y_2 +y_3)  \delta_{2L} (y_3 + y_1) \end{matrix} ~ \bigg\} \cr
\noalign{\kern 3pt}
&\quad
\Rightarrow \frac{1}{2} \Big\{ \delta (y_1) \delta (y_2) \delta (y_3) + \delta (y_1 -L) \delta (y_2-L) \delta (y_3 -L) \Big\}  ~, \cr
\noalign{\kern 10pt}
&\begin{matrix} \hat \delta_{2L} (y_1 - y_2)  \hat \delta_{2L} (y_2 - y_3)  \hat \delta_{2L} (y_3 + y_1) \cr
\noalign{\kern 5pt}
\hat \delta_{2L} (y_1 + y_2)  \hat \delta_{2L} (y_2 +y_3)  \hat \delta_{2L} (y_3 + y_1) \end{matrix} ~ \bigg\} \cr
\noalign{\kern 3pt}
&\quad 
\Rightarrow \frac{1}{2} \Big\{ \delta (y_1) \delta (y_2) \delta (y_3) - \delta (y_1 -L) \delta (y_2-L) \delta (y_3 -L) \Big\}  ~.
\label{RSAnomaly3}
\end{align}
As $h_n (0) = h_n (L) = 0$, only the terms proportional to $k_1 k_2 k_3$ in (\ref{RSAnomaly2}) survive.
We find the formula for the anomaly coefficients;
\begin{align}
a_{n \ell m} (\theta_H, z_L) &=  Q_0 k_n (0) k_\ell (0) k_m (0) + Q_1 k_n (L) k_\ell (L) k_m (L)  ~, \cr
\noalign{\kern 5pt}
(Q_0, Q_1) &= \begin{cases} (+1, +1) &\hbox{for type 1A} \cr   (-1, -1) &\hbox{for type 1B} \cr
 (+1, -1) &\hbox{for type 2A} \cr  (-1, +1) &\hbox{for type 2B} \end{cases} ~.
\label{RSAnomaly4}
\end{align}
The anomaly coefficients are determined by the values of the wave functions of the gauge fields at the UV and IR branes
and the parity conditions of the fermion fields.

\begin{figure}[tbh]
\centering
\includegraphics[width=65mm]{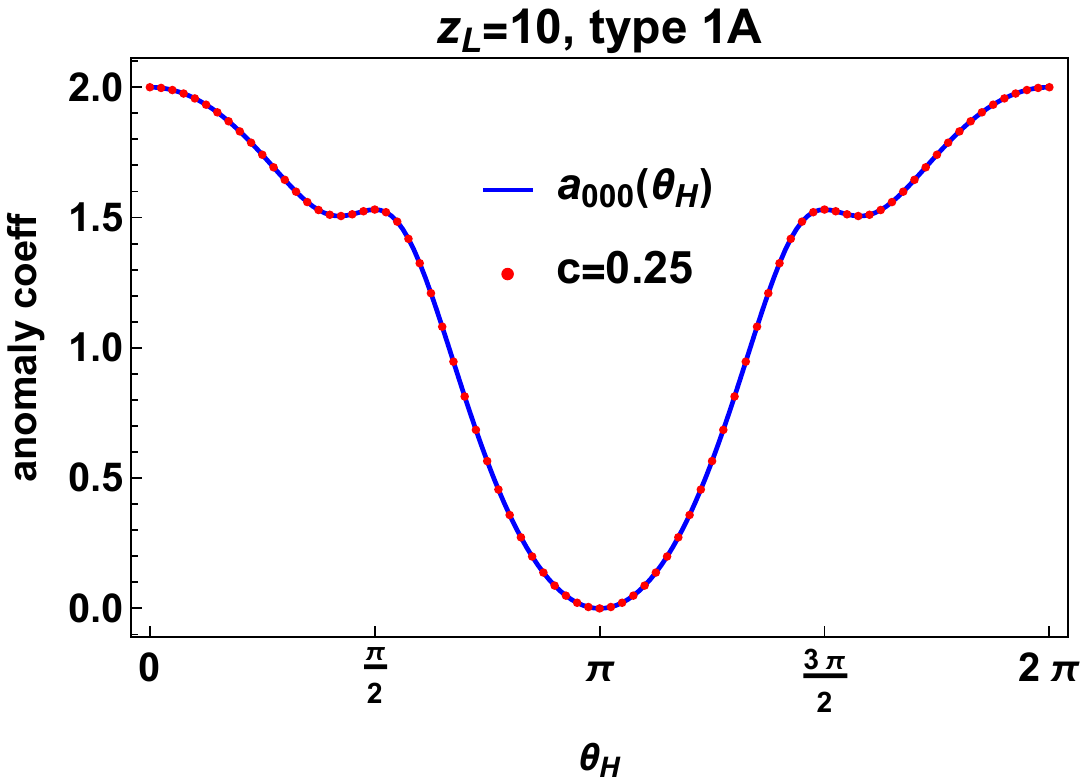}
~
\includegraphics[width=65mm]{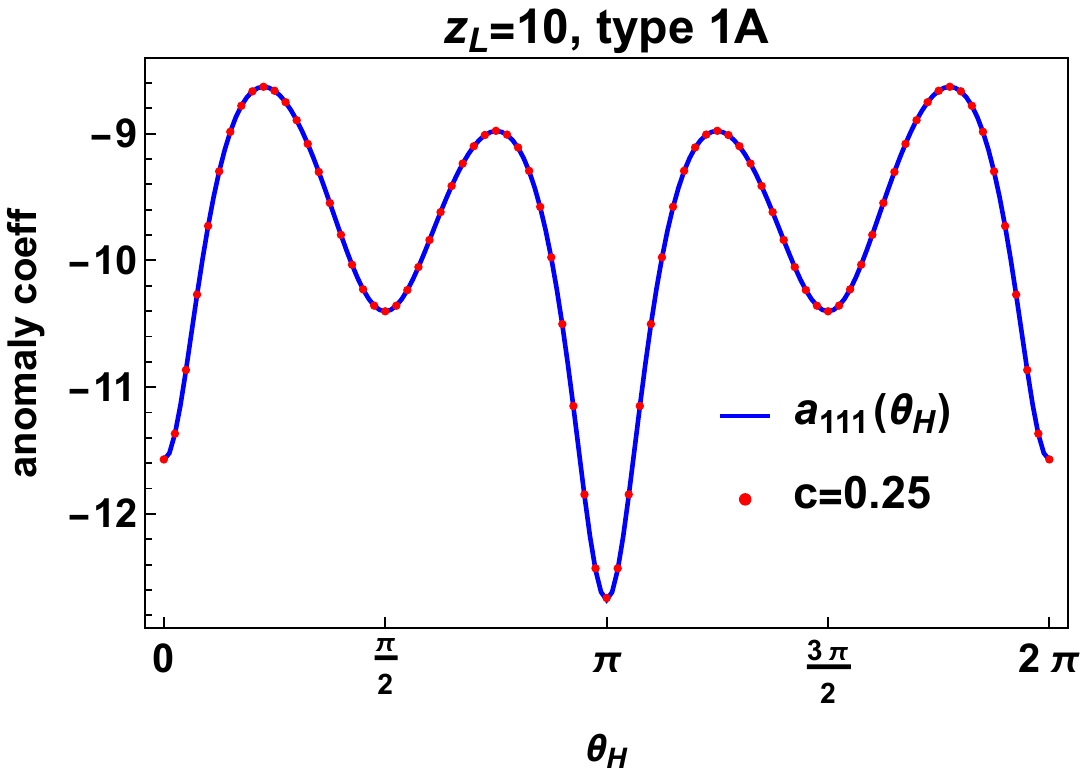}\\
\includegraphics[width=65mm]{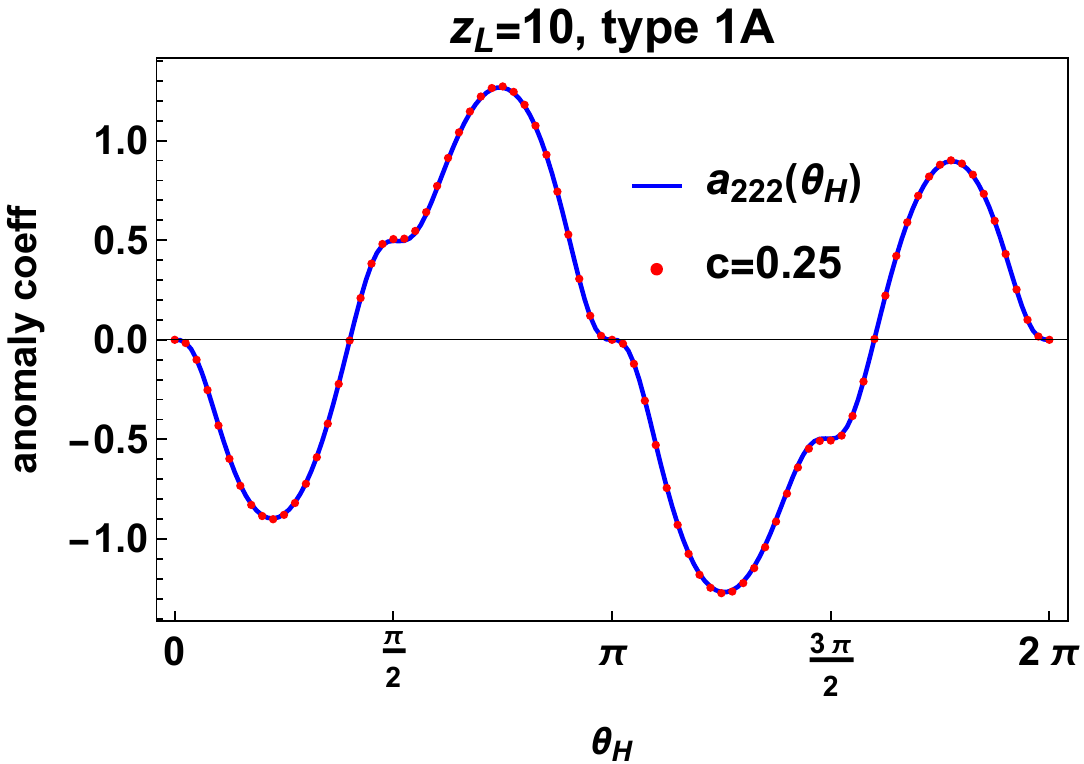}
~
\includegraphics[width=65mm]{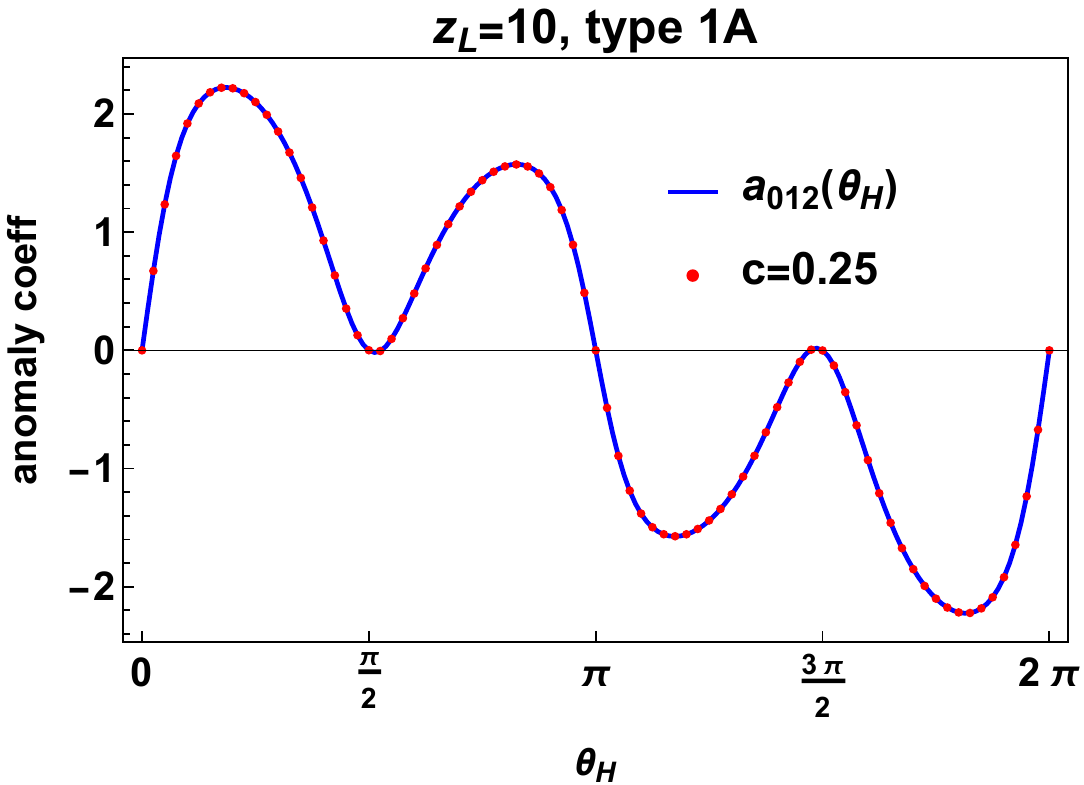}
\caption{The anomaly coefficients $a_{000}, a_{111}, a_{222}$  and $a_{012}$ as functions of $\theta_H$ 
are shown for type 1A fermions for $z_L=10$.
Blue curves represent the universal curves given by (\ref{RSAnomaly4}).
Red dots represent the values determined from the gauge couplings $t^{R/L}_{n\ell m}$ ($0 \le \ell, m \le \ell_0$) 
in (\ref{RSgaugeCoupling2}) and then taking the traces of $(\ell_0+1)$-dimensional matrices in (\ref{RSAnomaly1})
for fermions with $c=0.25$ and $\ell_0=10$.
}   
\label{fig:c025anomaly1}
\end{figure}

The formula (\ref{RSAnomaly4}) is strikingly simple.  The wave function $k_n(y)$ depends on $\theta_H$ and $z_L$.
The sum of the chiral anomalies arising from all possible fermion KK modes are summarized in terms of $k_n(0)$ and $k_n(L)$.
The $c$-independence of those anomalies is confirmed numerically.
The anomaly coefficients $a_{n\ell m}$  given by (\ref{RSAnomaly4}) are compared
with those determined by first evaluating the gauge couplings $t^{R/L}_{n\ell m}$ ($0 \le \ell, m \le \ell_0$) 
in (\ref{RSgaugeCoupling2}) and then taking the traces of $(\ell_0 +1)$-dimensional matrices in (\ref{RSAnomaly1}).
In Figure \ref{fig:c025anomaly1} the results for $a_{000}, a_{111}, a_{222}$  and $a_{012}$ are shown for type 1A fermions 
with $c=0.25$,  $\ell_0 =10$  and $z_L=10$.  
One sees that the numerically evaluated values for $c=0.25$ fall on the universal
curves given by (\ref{RSAnomaly4}).  
We have checked that the numerically evaluated values for other values of $c$ fall on the universal
curves as well.

\begin{figure}[tbh]
\centering
\includegraphics[width=65mm]{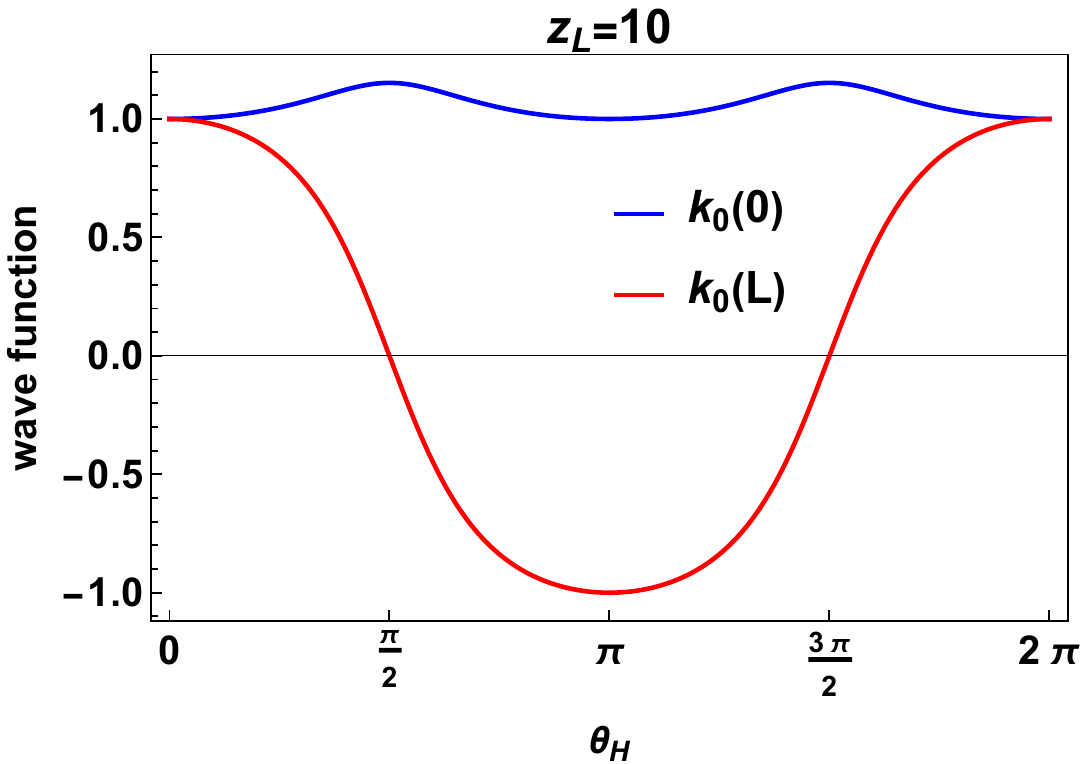}
~
\includegraphics[width=65mm]{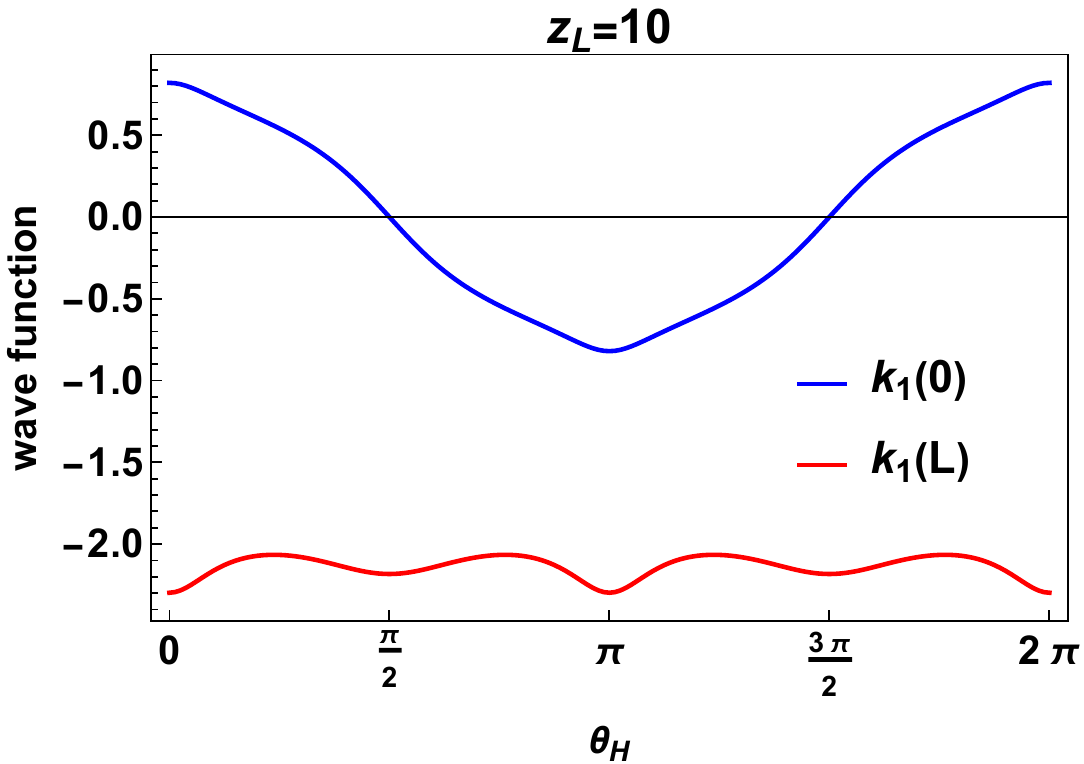}\\
\vskip 10pt
\includegraphics[width=65mm]{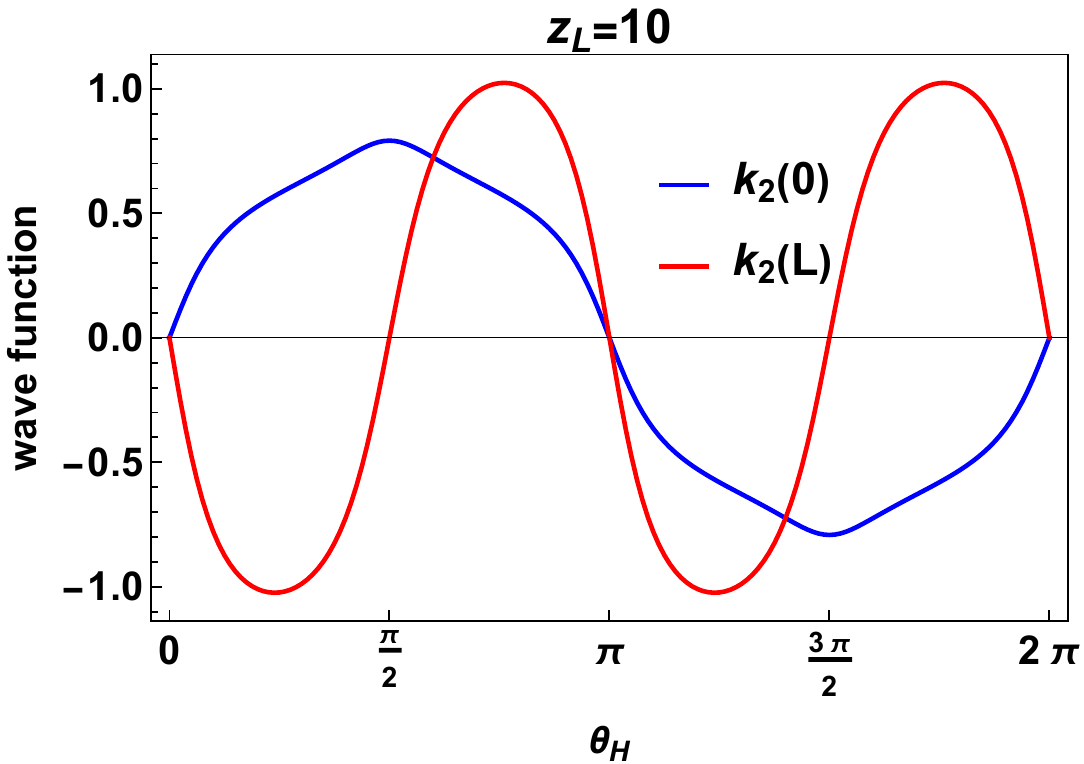}
~
\includegraphics[width=65mm]{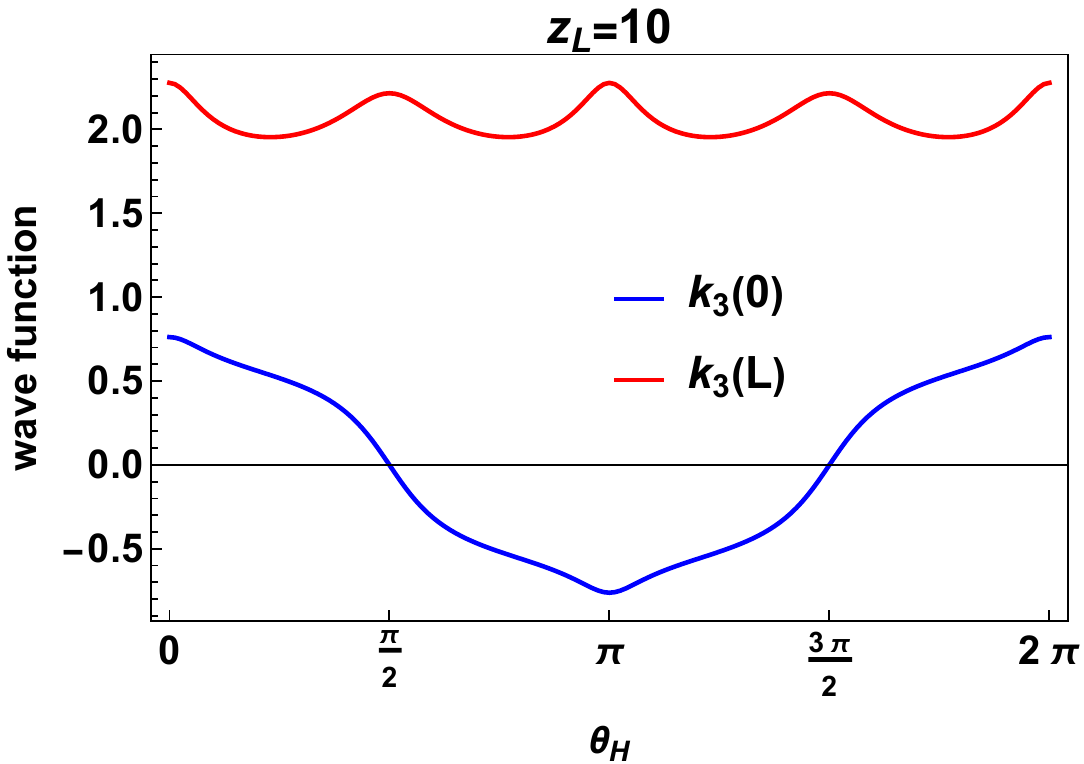}
\caption{The values of the gauge wave functions $ k_n(y; \theta_H)$ ($n=0,1,2,3$) at $y=0$ (blue curves) and 
$y=L$ (red curves) for $z_L=10$ are shown.
}   
\label{fig:waveFunction1}
\end{figure}

Some of $ k_n(0; \theta_H)$ and $k_n(L; \theta_H)$ are plotted in Figure \ref{fig:waveFunction1}.
Note that for $n=1,3,5, \cdots$, $|k_n (L; \theta_H) | $ is much larger than $|k_n (0; \theta_H) | $ for $z_L \ge 10$.
Massless gauge bosons ($Z_\mu^{(0)}$) exist at $\theta_H = 0$ and $\pi$.
$k_0(0; 0) = k_0(L; 0) = 1$ and $k_0(0; \pi) = -  k_0(L; \pi) = 1$ so that
$a_{000} (\theta_H=0) = 2$ and $a_{000} (\theta_H=\pi) = 0$ for type 1A fermions and
$a_{000} (\theta_H=0) = 0$ and $a_{000} (\theta_H=\pi) = 2$ for type 2A fermions.
The anomaly flow is reflected in the behavior of the wave functions of the gauge fields at $y=0$ and $L$.

\begin{figure}[tbh]
\centering
\includegraphics[width=65mm]{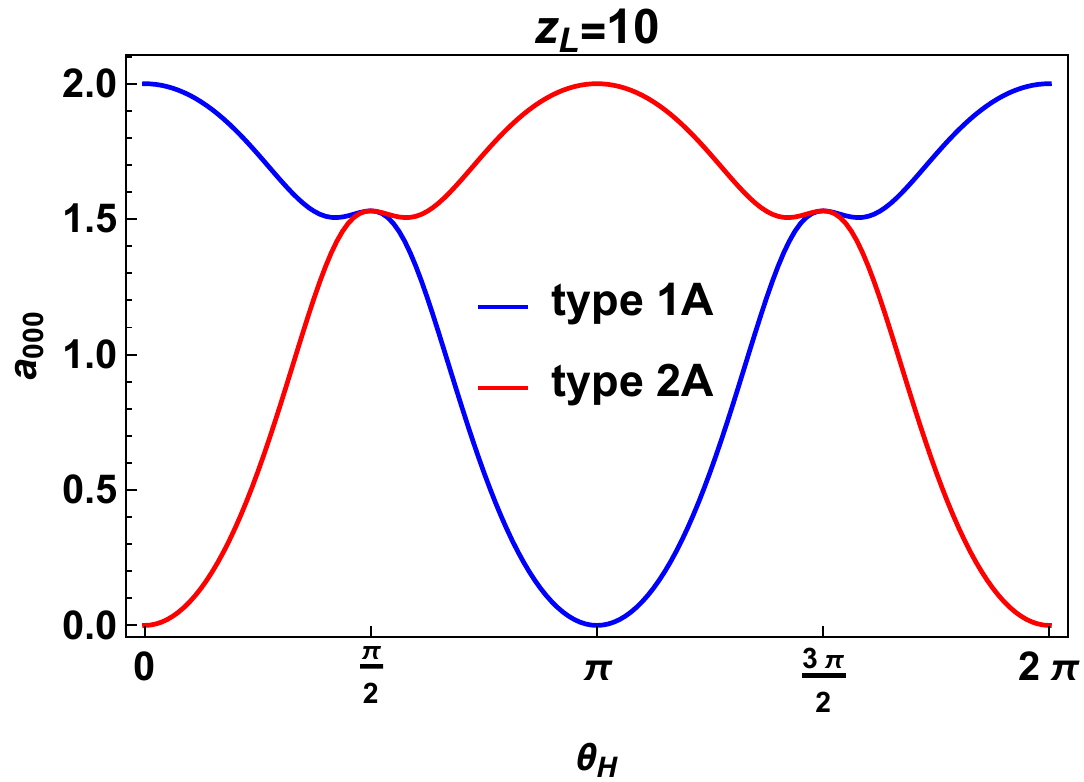}
~
\includegraphics[width=65mm]{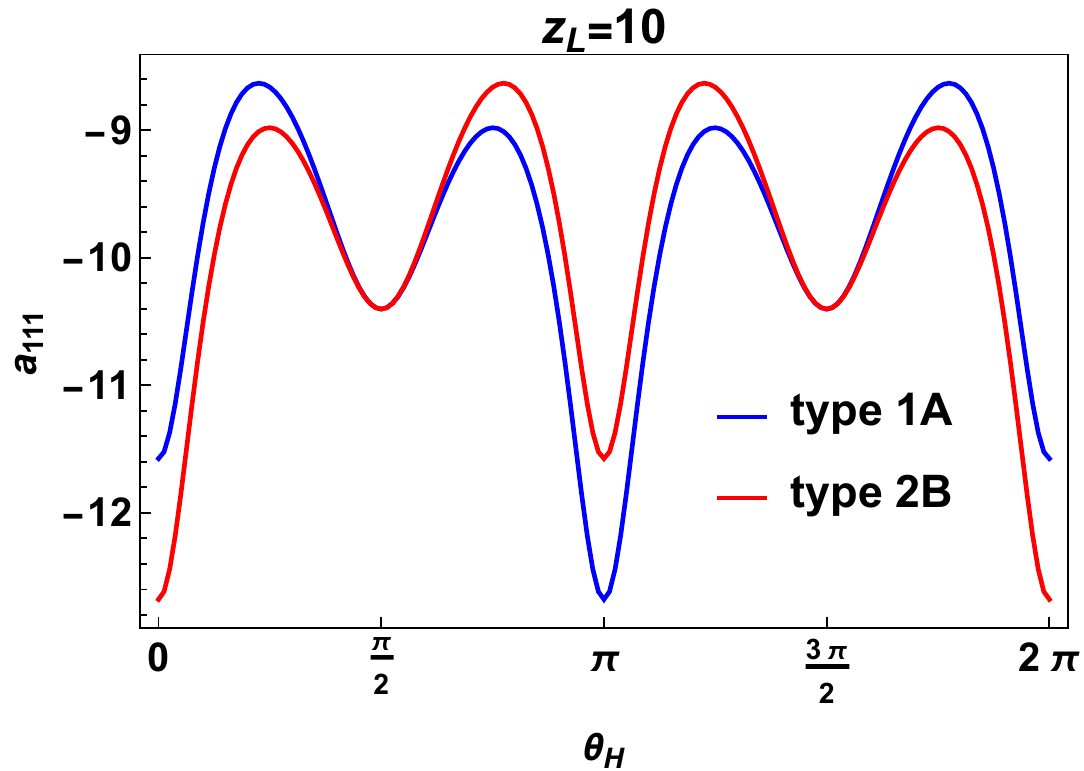}\\
\vskip 10pt
\includegraphics[width=65mm]{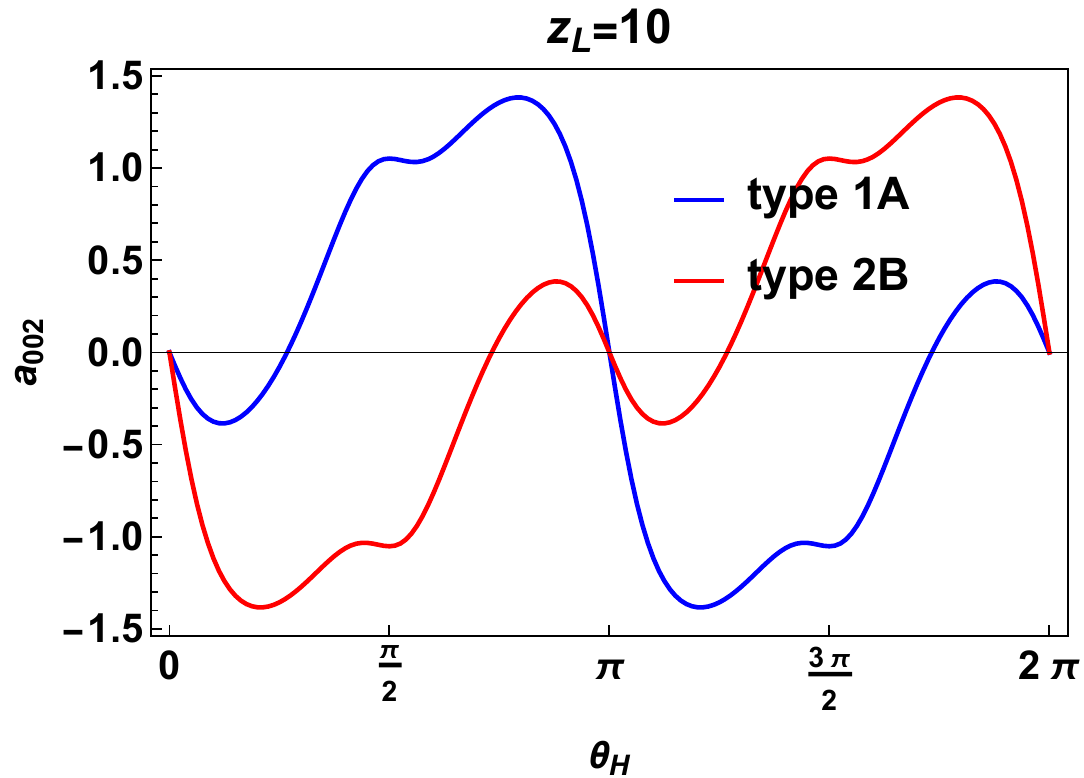}
~
\includegraphics[width=65mm]{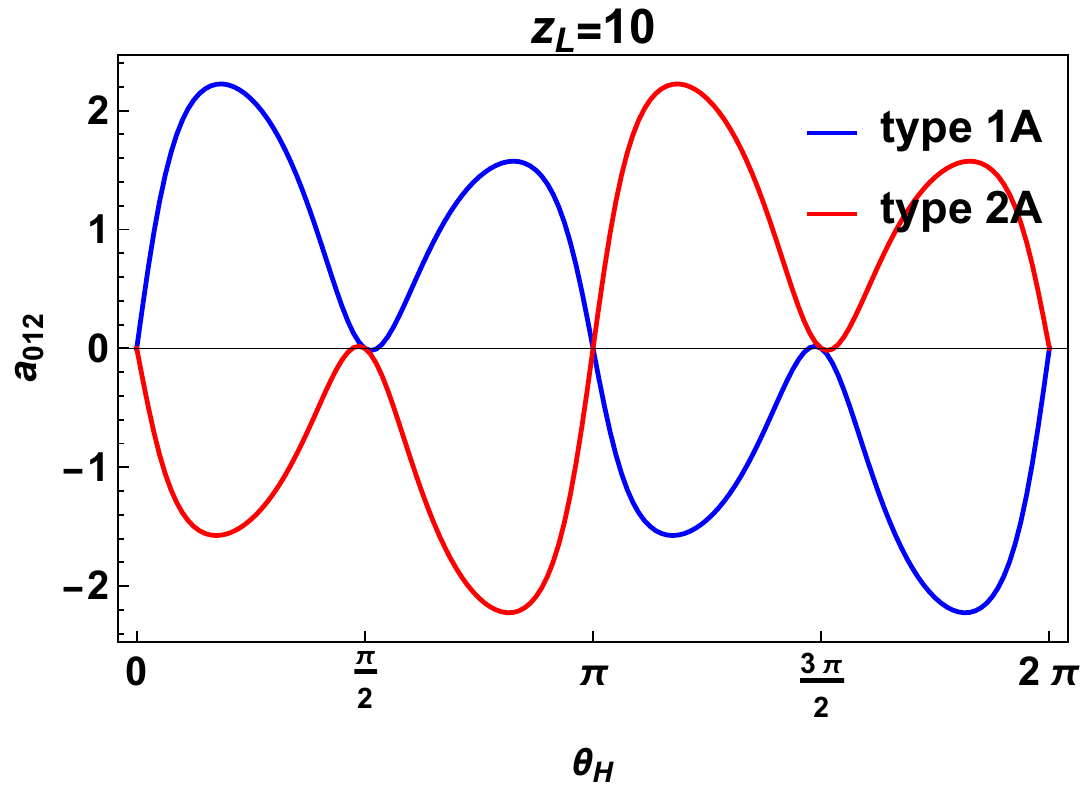}
\caption{Dependence of the anomaly coefficients  $a_{000}, a_{111}, a_{002}$ and $a_{012}$
on fermion types is shown for $z_L=10$.
One sees that $a_{n\ell m}(\theta_H + \pi)^{\rm{type}\, \rm{1A}} = a_{n\ell m}(\theta_H)^{\rm{type}\, \rm{2A}}$ or
$a_{n\ell m}(\theta_H)^{\rm{type}\, \rm{2B}}$.
}   
\label{fig:anomaly-type}
\end{figure}

Dependence of the anomaly coefficients $a_{n\ell m}$ on fermion types has a simple pattern.
$a_{n\ell m}(\theta_H)^{\rm{type}\, \rm{1A}} = - a_{n\ell m}(\theta_H)^{\rm{type}\, \rm{1B}}$ and
$a_{n\ell m}(\theta_H)^{\rm{type}\, \rm{2A}} = - a_{n\ell m}(\theta_H)^{\rm{type}\, \rm{2B}}$.
Further $a_{n\ell m}(\theta_H + \pi)^{\rm{type}\, \rm{1A}} = a_{n\ell m}(\theta_H)^{\rm{type}\, \rm{2A}}$ or
$a_{n\ell m}(\theta_H)^{\rm{type}\, \rm{2B}}$. (See Figure \ref{fig:anomaly-type}.)  It follows from the property that 
$[k_n(0) , k_n(L)]_{\theta_H + \pi} = [k_n(0) , - k_n(L)]_{\theta_H}$ or $ [-k_n(0) , k_n(L)]_{\theta_H}$.

Formulas in the flat $M^4 \times (S^1/Z_2)$ spacetime simplify.
With the KK expansions (\ref{gaugeKKflat2}), (\ref{fermionKKflat2}), and (\ref{fermionKKflat3}), 
the gauge couplings are written as
\begin{align}
&\frac{g_4}{2} \sum_{n=-\infty}^\infty \sum_{\ell = -\infty}^\infty \sum_{m=-\infty}^\infty B_\mu^{(n)} (x)
\Big\{ s^R_{n\ell m} \, \psi_R^{(\ell)} (x)^\dagger \bar \sigma^\mu \psi_R^{(m)} (x) 
+ s^L_{n\ell m} \, \psi_L^{(\ell)} (x)^\dagger  \sigma^\mu \psi_L^{(m)} (x) \Big\} 
\label{flatGaugeCoupling1}
\end{align}
for type 1A and 1B fermions.  For type 2A and 2B fermions $\psi_{R/L}^{(m)} (x) $ should be replaced 
by $\psi_{R/L}^{(m + \frac{1}{2})} (x) $. 
The anomaly coefficient associated with the three legs  of 
$B_{\mu_1}^{(n_1)} B_{\mu_2}^{(n_2)} B_{\mu_3}^{(n_3)}$ is given by 
\begin{align}
&b_{n_{1} n_{2} n_{3}} = b_{n_{1} n_{2} n_{3}}^{R} + b_{n_{1} n_{2} n_{3}}^{L} ~, \cr
\noalign{\kern 5pt}
&b_{n_{1} n_{2} n_{3}}^{R} = \Tr S^{R}_{n_{1}} S^{R}_{n_{2}} S^{R}_{n_{3}} ~,~~
(S^{R}_{n} )_{m \ell} = s^{R}_{n m \ell} ~, \cr
\noalign{\kern 5pt}
&b_{n_{1} n_{2} n_{3}}^{L} = \Tr S^{L}_{n_{1}} S^{L}_{n_{2}} S^{L}_{n_{3}} ~,~~
(S^{L}_{n} )_{m \ell} = s^{L}_{n m \ell} ~.
\label{flatAnomaly1}
\end{align}
Applying the same argument as in the case of the RS space, one finds that
\begin{align}
b_{n \ell m}  &=  Q_0 k^{\rm flat}_n (0) k^{\rm flat}_\ell (0) k^{\rm flat}_m (0) 
+ Q_1 k^{\rm flat}_n (L) k^{\rm flat}_\ell (L) k^{\rm flat}_m (L)  
\label{flatAnomaly2}
\end{align}
where $Q_0, Q_1$ are given in (\ref{RSAnomaly4}).  
Since $k^{\rm flat}_n(y) = \cos (n\pi y/L)$ from  (\ref{gaugeKKflat2}), one finds that
\begin{align}
b_{n \ell m}  &=  Q_0 + (-1)^{n + \ell + m}  \,  Q_1 ~,
\label{flatAnomaly3}
\end{align}
which agrees with the result in Ref.\cite{anomalyFlow2022}.
The formula (\ref{flatAnomaly3}) also results in the flat spacetime limit of (\ref{RSAnomaly4}).
In the flat spacetime the level-crossing in the mass spectrum of gauge fields occurs at $\theta_H=0, \pm \onehalf \pi,
\pm \pi, \cdots$.  For this reason the flat spacetime limit of  (\ref{RSAnomaly4}) becomes singular, as has been
shown in Ref.\cite{anomalyFlow2022}.

\section{Anomaly cancellation}

The universality of the anomaly flow, expressed in the formula (\ref{RSAnomaly4}), has a profound implication
in the model building, particularly in the GHU scenario.
Chiral anomalies associated with gauge currents must be cancelled for the consistency of the  theory 
in four dimensions \cite{Bouchiat1972, GrossJackiw1972}.
The fact that the anomaly coefficients are independent of the bulk mass parameters  of fermions implies that
anomaly cancellation can be achieved among various distinct fermions in the theory.
In this section we examine this problem in the $SU(2)$ model.

Let us first recall the equations following from the action $I_\RS$ in (\ref{RSaction}) are, at the classical level, 
\begin{align}
&\frac{1}{\sqrt{ - \det G}} \, \dd_M \big( \sqrt{ - \det G} \, F^{MN} \big) - ig_A [A_M, F^{MN}] + J^N =0 ~, \cr
\noalign{\kern 5pt}
&{\cal D} (c) \, \Psi = 0 ~,\cr
\noalign{\kern 5pt}
&J^N = J^{Na}  \frac{\tau^a}{2} ~, ~~
J^{Na} = - i g_A \overline{\Psi} \gamma^A {e_A}^N \frac{\tau^a}{2} \Psi  ~.
\label{EoM1}
\end{align}
The current in five dimensions is covariantly conserved;
\begin{align}
&\frac{1}{\sqrt{ - \det G}} \, \dd_N \big( \sqrt{ - \det G} \, J^N \big) -  ig_A [A_N, J^N] = 0 ~.
\label{EoM2}
\end{align}
Note that the derivative term in the fifth coordinate generates mass terms in four dimensions when 
expanded in the KK modes.
At the quantum level there arises an anomaly term on the righthand side of Eq.\ (\ref{EoM2}). 
The four-dimensional current $j_{(n)}^{\mu} (x)$ which couples with $Z_\mu^{(n)} (x)$ is
\begin{align}
j_{(n)}^{\mu} (x) &= \int_0^L dy \, \sqrt{ - \det G} \,  \big\{ h_n (y) J^{\mu 1} + k_n(y) J^{\mu 3} \big\} \cr
\noalign{\kern 5pt} 
&= 
\frac{g_4}{2} \sum_{\ell = 0}^\infty \sum_{m=0}^\infty 
\Big\{ t^R_{n\ell m} \, \chi_R^{(\ell)} (x)^\dagger \bar \sigma^\mu \chi_R^{(m)} (x) 
+ t^L_{n\ell m} \, \chi_L^{(\ell)} (x)^\dagger  \sigma^\mu \chi_L^{(m)} (x) \Big\} .
\label{current1}
\end{align}
The divergence $\dd_\mu j_{(n)}^{\mu}$ picks up an anomalous term $j_{(n)}^{\rm anomaly}$ given by
\begin{align}
j_{(n)}^{\rm anomaly} = - \Big( \frac{g_4}{2} \Big)^3
\sum_{\ell = 0}^\infty \sum_{m=0}^\infty  \frac{a_{n\ell m}}{32 \pi^2} \,  \ep^{\mu\nu\rho\sigma}
Z^{(\ell)}_{\mu \nu}  Z^{(m)}_{\rho\sigma} 
\label{current2}
\end{align}
where $Z^{(\ell)}_{\mu \nu} = \dd_\mu Z_\nu^{(\ell)} - \dd_\nu Z_\mu^{(\ell)}$.

The conditions for the cancellation of the gauge anomalies are simple.  Let the numbers of doublet fermions
of type 1A, 1B, 2A and 2B be $n_{1A}$, $n_{1B}$, $n_{2A}$ and $n_{2B}$, respectively.  
It follows from (\ref{RSAnomaly4}) that the anomalies are cancelled  if 
\begin{align}
&n_{1A} = n_{1B}  ~,~~ n_{2A} = n_{2B}  ~.
\label{cancellation1}
\end{align}
In the presence of brane fermions, namely fermions living only on the UV or IR brane, the conditions 
are generalized.  Suppose that there are $\hat n_R$ right-handed and  $\hat n_L$ left-handed  doublet brane fermions 
on the UV brane at $y=0$.
As the $Z_\mu^{(n)}$ coupling of each brane fermion is given by $(g_4/2) \, k_n(0)$, the anomaly
cancellation conditions become
\begin{align}
&n_{1A} - n_{1B}  + n_{2A} - n_{2B}  + \hat n_R - \hat n_L = 0 ~, \cr
&n_{1A} - n_{1B}  - n_{2A} + n_{2B} = 0 ~,
\label{cancellation2}
\end{align}
We stress that the conditions (\ref{cancellation1}) and (\ref{cancellation2}) do not depend on $\theta_H$ and $z_L$.
Furthermore the conditions guarantee that not only the zero mode anomaly $a_{000}$ but also all other anomalies 
$a_{n\ell m}$ are cancelled at once.

Fermion multiplets in the triplet representation do not contribute to anomalies in the $SU(2)$ gauge theory as is easily 
confirmed. The anomaly cancellation is achieved by the condition (\ref{cancellation1}) or (\ref{cancellation2}), namely
by the condition for the numbers of doublet fermions with four types of orbifold boundary conditions.  
It does not depend on the AB phase  $\theta_H$, namely the VEV of $A_y$.  
The situation is very similar to the anomaly cancellation condition in the SM.

\section{Summary and discussions}

In this paper we have examined the anomaly flow by the AB phase $\theta_H$ in the $SU(2)$ gauge theory 
in the RS space and in the flat  $M^4 \times (S^1/Z_2)$ spacetime.
The anomaly coefficients $a_{n\ell m} (\theta_H, z_L)$ induced by a fermion field  in the bulk smoothly changes 
in $\theta_H$ in the RS space.  Although the gauge couplings of the fermion, $t_{n\ell m}^{R/L} (\theta_H, z_L, c)$,
nontrivially depend on the bulk mass parameter $c$ of the fermion, the total anomaly coefficients $a_{n\ell m} $
are independent of $c$.  We have shown that those anomaly coefficients $a_{n\ell m} $ are expressed
in terms of the values of the wave functions of the gauge fields at the UV and IR branes.
The holographic formula (\ref{RSAnomaly4}) manifestly exhibits the $c$-independence.
We have confirmed that the values of the anomaly coefficients numerically evaluated directly from 
$t_{n\ell m}^{R/L} (\theta_H, z_L, c)$ fall precisely on the curves given by (\ref{RSAnomaly4}).
It has been left for future investigation to find an analytic proof of the $c$-independence of the 
expression (\ref{RSAnomaly2}).

As has been mentioned in the previous section, the universality in  anomaly flow is critically important
in the construction of realistic models of particle physics.  GHU models have been proposed to unify
the 4D Higgs boson with gauge fields in the framework of gauge theory on five-dimensional orbifolds 
in which the gauge hierarchy problem is naturally solved \cite{Hatanaka1998, Kubo2002, Burdman2003,
Csaki2003, Scrucca2003, Agashe2005, Cacciapaglia2006, Medina2007, HOOS2008, HNU2010, 
Amodel2013, GUTinspired2019, Yoon2019, FCNC2020a, GUTinspired2020c}. 
In particular, $SO(5) \times U(1)_X \times SU(3)_C$ GHU in the RS space with $\theta_H \sim 0.1$ and 
$z_L = 10^5 \sim 10^{10}$  has been shown to reproduce
nearly the same phenomenology at low energies as the SM \cite{GUTinspired2019, FCNC2020a}.
As in the case of the SM, all chiral anomalies associated with 
gauge currents must be cancelled.  Generalization of the argument on the universality  to the group 
$SO(5) \times U(1)_X \times SU(3)_C$ is necessary. 
Further the technology developed in the present paper can be applied to the evaluation of
anomalies of global currents such as baryon and lepton numbers.  
The phenomenon of anomaly flow may possibly be related to Chern-Simons terms 
in five dimensions\cite{Gripaios2008, Hong2021, Lim2021}.
These issues will be clarified in separate papers.

\section*{Acknowledgement}

This work was supported in part by Japan Society for the Promotion of Science, Grants-in-Aid for Scientific 
Research, Grant No. JP19K03873.

\vskip 1.5cm


\appendix

\section{Basis functions} 
Wave functions of gauge fields and fermions are expressed in terms of the following basis functions.
For gauge fields we introduce
\begin{align}
 C(z; \lambda) &= \frac{\pi}{2} \lambda z z_L F_{1,0}(\lambda z, \lambda z_L) ~,  \cr
 S(z; \lambda) &= -\frac{\pi}{2} \lambda  z F_{1,1}(\lambda z, \lambda z_L) ~, \cr
 C^\prime (z; \lambda) &= \frac{\pi}{2} \lambda^2 z z_L F_{0,0}(\lambda z, \lambda z_L) ~,  \cr
S^\prime (z; \lambda) &= -\frac{\pi}{2} \lambda^2 z  F_{0,1}(\lambda z, \lambda z_L)~, \cr
\noalign{\kern 5pt}
 F_{\alpha, \beta}(u, v) &\equiv 
J_\alpha(u) Y_\beta(v) - Y_\alpha(u) J_\beta(v) ~,
\label{functionA1}
\end{align}
where $J_\alpha (u)$ and $Y_\alpha (u)$ are Bessel functions of  the first and second kind.
They satisfy
\begin{align}
&- z \frac{d}{dz} \frac{1}{z} \frac{d}{dz} \begin{pmatrix} C \cr S \end{pmatrix} 
= \lambda^{2} \begin{pmatrix} C \cr S \end{pmatrix} ~, \cr
\noalign{\kern 5pt}
&C(z_{L} ; \lambda)  = z_{L} ~, ~~ C' (z_{L} ; \lambda)  = 0 ~, \cr
\noalign{\kern 5pt}
&S(z_{L} ; \lambda)  = 0 ~, ~~ S' (z_{L} ; \lambda)  = \lambda ~,  \cr
\noalign{\kern 5pt}
&CS' - S C' = \lambda z ~.
\label{relationA1}
\end{align}
To express wave functions of KK modes of gauge fields,  we  make use of
\begin{align}
&\hat S(z;\lambda) = N_{0}(\lambda) S(z;\lambda)  ~,~~
\hat C(z;\lambda) = N_{0}(\lambda)^{-1} C(z;\lambda)  ~, \cr
\noalign{\kern 5pt}
&\check S(z;\lambda) = N_{1}(\lambda) S(z;\lambda)  ~,~~
\check C(z;\lambda) = N_{1}(\lambda)^{-1} C(z;\lambda)  ~, \cr
\noalign{\kern 5pt}
&\qquad N_{0}(\lambda) = \frac{C(1;\lambda)}{S(1;\lambda)} ~, ~~
N_{1}(\lambda) = \frac{C'(1;\lambda)}{S'(1;\lambda)} ~.
\label{functionA2}
\end{align}

For fermion fields with a bulk mass parameter $c$, we define 
\begin{align}
\begin{pmatrix} C_L \cr S_L \end{pmatrix} (z; \lambda,c)
&= \pm \frac{\pi}{2} \lambda \sqrt{z z_L} F_{c+\frac12, c\mp\frac12}(\lambda z, \lambda z_L) ~, \cr
\begin{pmatrix} C_R \cr S_R \end{pmatrix} (z; \lambda,c)
&= \mp \frac{\pi}{2} \lambda \sqrt{z z_L} F_{c- \frac12, c\pm\frac12}(\lambda z, \lambda z_L) ~.
\label{functionA3}
\end{align}
These functions satisfy 
\begin{align}
&D_{+} (c) \begin{pmatrix} C_{L} \cr S_{L} \end{pmatrix} = \lambda  \begin{pmatrix} S_{R} \cr C_{R} \end{pmatrix}, \cr
\noalign{\kern 5pt}
&D_{-} (c) \begin{pmatrix} C_{R} \cr S_{R} \end{pmatrix} = \lambda  \begin{pmatrix} S_{L} \cr C_{L} \end{pmatrix}, ~~
D_{\pm} (c) = \pm \frac{d}{dz} + \frac{c}{z} ~, \cr
\noalign{\kern 5pt}
&C_{R} = C_{L} = 1 ~, ~~ S_{R} = S_{L} = 0 \quad {\rm at~} z=z_{L} ~, \cr
\noalign{\kern 5pt}
&C_L C_R - S_L S_R=1 ~. 
\label{relationA2}
\end{align}
Also $C_L  (z; \lambda, -c) = C_R  (z; \lambda, c)$ and $S_L  (z; \lambda, -c) = - S_R  (z; \lambda, c)$ hold.  
To express wave functions of KK modes of fermion fields,  we  make use of
\begin{align}
&\hat S_{L}  (z; \lambda,c) = N_{L}( \lambda,c) S_{L}  (z; \lambda,c) ~, ~~
\hat C_{L}  (z; \lambda,c) = N_{R}( \lambda,c) C_{L}  (z; \lambda,c) ~,  \cr
\noalign{\kern 5pt}
&\hat S_{R}  (z; \lambda,c) = N_{R}( \lambda,c) S_{R}  (z; \lambda,c) ~, ~~
\hat C_{R}  (z; \lambda,c) = N_{L}( \lambda,c) C_{R}  (z; \lambda,c) ~,  \cr
\noalign{\kern 5pt}
&\check S_{L}  (z; \lambda,c) = N_{R}( \lambda,c)^{-1} S_{L}  (z; \lambda,c) ~, ~~
\check C_{L}  (z; \lambda,c) = N_{L}( \lambda,c)^{-1} C_{L}  (z; \lambda,c) ~,  \cr
\noalign{\kern 5pt}
&\check S_{R}  (z; \lambda,c) = N_{L}( \lambda,c)^{-1} S_{R}  (z; \lambda,c) ~, ~~
\check C_{R}  (z; \lambda,c) = N_{R}( \lambda,c)^{-1} C_{R}  (z; \lambda,c) ~,  \cr
\noalign{\kern 5pt}
&\qquad N_{L}( \lambda,c)  = \frac{C_{L} (1; \lambda,c)}{S_{L} (1; \lambda,c)} ~, ~~
N_{R}( \lambda,c)  = \frac{C_{R} (1; \lambda,c)}{S_{R} (1; \lambda,c)} ~.
\label{functionA4}
\end{align}

\section{Wave functions in RS} 

\subsection{Gauge fields $Z_\mu^{(n)}$}

The mode functions of the gauge fields $Z_\mu^{(n)} (x)$ in (\ref{RSgaugeKK1}) are given by
\begin{align}
 \tilde{\bf h}_0 (z) &= \bar {\bf h}_0^a (z) ~, \cr
\noalign{\kern 5pt}
 \tilde{\bf h}_{2\ell - 1} (z) &= (-1)^\ell \begin{cases} 
\bar {\bf h}_{2\ell - 1}^a (z) &{\rm for} - \frac{1}{2} \pi  <  \theta_H < \frac{1}{2} \pi \cr
\bar {\bf h}_{2\ell - 1}^b (z) &{\rm for~} 0 < \theta_H < \pi \cr
- \bar {\bf h}_{2\ell - 1}^a (z) &{\rm for~}  \frac{1}{2} \pi < \theta_H < \frac{3}{2} \pi \cr
-\bar {\bf h}_{2\ell - 1}^b (z) &{\rm for~}  \pi < \theta_H < 2 \pi \cr
\bar {\bf h}_{2\ell - 1}^a (z) &{\rm for~} \frac{3}{2} \pi <  \theta_H <  \frac{5}{2} \pi 
\end{cases} ~  ~(\ell = 1, 2, 3, \cdots), 
\cr
\noalign{\kern 5pt}
 \tilde{\bf h}_{2\ell} (z) &= (-1)^\ell \begin{cases} 
\bar {\bf h}_{2\ell }^b (z) &{\rm for} - \frac{1}{2} \pi  <  \theta_H < \frac{1}{2} \pi \cr
- \bar {\bf h}_{2\ell}^a (z) &{\rm for~} 0 <  \theta_H  < \pi \cr
- \bar {\bf h}_{2\ell}^b (z) &{\rm for~}  \frac{1}{2} \pi < \theta_H < \frac{3}{2} \pi \cr
\bar {\bf h}_{2\ell}^a (z) &{\rm for~}  \pi < \theta_H <  2 \pi \cr
\bar {\bf h}_{2\ell }^b (z) &{\rm for~} \frac{3}{2} \pi <  \theta_H < \frac{5}{2} \pi 
\end{cases}  ~(\ell = 1, 2, 3, \cdots), 
\cr
\noalign{\kern 10pt}
\bar {\bf h}_n^a (z) 
&= \frac{1}{\sqrt{r^a_n}} \begin{pmatrix} - s_{H}  \hat S (z; \lambda_n) \cr c_{H} C (z; \lambda_n)   \end{pmatrix} , ~~
\bar {\bf h}_n^b (z)  
= \frac{1}{\sqrt{r^b_n}} \begin{pmatrix} c_{H}   S (z; \lambda_n) \cr s_{H} \check C (z; \lambda_n)   \end{pmatrix} , \cr
\noalign{\kern 5pt}
&s_H = \sin \theta_H ~, ~~ c_H = \cos \theta_H ~, \cr
\noalign{\kern 5pt}
&r_{n} = \frac{1}{kL} \int_{1}^{z_{L}} \frac{dz}{z} \big\{ | \hat h_{n} (z) |^{2} + | \hat k_{n} (z) |^{2} \big\}
\quad {\rm for}~ \begin{pmatrix} \hat h_{n} (z) \cr \hat k_{n} (z) \end{pmatrix} .
\label{RSgaugeKKB1}
\end{align}
$\hat S$ and $\check C$ are given in (\ref{functionA2}).
In the above  formulas, the two expressions given in an overlapping $\theta_H$ region  are the same.
The connection formulas are necessary as one of them fails to make sense at the boundary in $\theta_H$.

\subsection{Fermion fields $\chi_{R/L}^{(n)}$}

The mode functions of the fermion fields $\chi_{R/L}^{(n)} (x)$ in (\ref{RSfermionKK1}) are given, 
for type 1A and $c>0$,  by
\begin{align}
\underline{\hbox{type 1A}} & \cr
\tilde{\bf f}_{R, 2\ell} (z)  &= 
\begin{cases} \bar {\bf f}_{R, 2\ell}^a (z) &{\rm for~} - \pi < \theta_H < \pi  \cr
\bar {\bf f}_{R, 2\ell}^b (z) &{\rm for~} 0 < \theta_H <  2\pi \cr
- \bar {\bf f}_{R, 2\ell}^a (z) &{\rm for~}   \pi < \theta_H < 3\pi  \cr
- \bar {\bf f}_{R, 2\ell}^b (z) &{\rm for~} 2 \pi < \theta_H <  4\pi \cr
\bar {\bf f}_{R, 2\ell}^a (z) &{\rm for~}  3 \pi < \theta_H < 5 \pi  \cr \end{cases} ~ (\ell=0,1,2, \cdots),\cr
\noalign{\kern 5pt}
\tilde{\bf f}_{R, 2\ell-1} (z)  &=  
\begin{cases} \bar {\bf f}_{R, 2\ell-1}^c (z) &{\rm for~} - \pi < \theta_H < \pi  \cr
\bar {\bf f}_{R, 2\ell-1}^d (z) &{\rm for~} 0 < \theta_H <  2\pi \cr
- \bar {\bf f}_{R, 2\ell-1}^c (z) &{\rm for~}   \pi < \theta_H < 3\pi  \cr
- \bar {\bf f}_{R, 2\ell-1}^d (z) &{\rm for~} 2 \pi < \theta_H <  4\pi \cr
\bar {\bf f}_{R, 2\ell-1}^c (z) &{\rm for~}  3 \pi < \theta_H < 5 \pi  \cr \end{cases} ~ (\ell=1,2,3, \cdots) , \cr
\noalign{\kern 10pt}
\tilde{\bf f}_{L0} (z)  &=  \bar {\bf f}_{L0}^a (z) , \cr
\noalign{\kern 5pt}
\tilde{\bf f}_{L, 2\ell-1} (z)  &= 
\begin{cases} \bar {\bf f}_{L, 2\ell-1}^a (z) &{\rm for~} - \pi < \theta_H < \pi  \cr
\bar {\bf f}_{L, 2\ell-1}^b (z) &{\rm for~} 0 < \theta_H <  2\pi \cr
- \bar {\bf f}_{L, 2\ell-1}^a (z) &{\rm for~}   \pi < \theta_H < 3\pi  \cr
- \bar {\bf f}_{L, 2\ell-1}^b (z) &{\rm for~} 2 \pi < \theta_H <  4\pi \cr
\bar {\bf f}_{L, 2\ell-1}^a (z) &{\rm for~}  3 \pi < \theta_H < 5 \pi  \cr \end{cases} ~ (\ell=1,2,3, \cdots) , \cr
\noalign{\kern 5pt}
\tilde{\bf f}_{L, 2\ell} (z)  &= 
\begin{cases} \bar {\bf f}_{L, 2\ell}^c (z) &{\rm for~} - \pi < \theta_H < \pi  \cr
\bar {\bf f}_{L, 2\ell}^d (z) &{\rm for~} 0 < \theta_H <  2\pi \cr
- \bar {\bf f}_{L, 2\ell}^c (z) &{\rm for~}   \pi < \theta_H < 3\pi  \cr
- \bar {\bf f}_{L, 2\ell}^d (z) &{\rm for~} 2 \pi < \theta_H <  4\pi \cr
\bar {\bf f}_{L, 2\ell}^c (z) &{\rm for~}  3 \pi < \theta_H < 5 \pi  \cr \end{cases} ~ (\ell=1,2, 3,\cdots),
\label{RSfermionKKB1}
\end{align}
Here
\begin{align}
&\bar {\bf f}_{Rn}^a (z) = \frac{1}{\sqrt{r^a_n}} 
\begin{pmatrix}  \bar c_{H} C_{R}(z; \lambda_{n}, c) \cr - \bar s_{H} \hat S_{R} (z; \lambda_{n}, c)  \end{pmatrix} , ~~
\bar {\bf f}_{Rn}^b (z)  = \frac{1}{\sqrt{r^b_n}} 
\begin{pmatrix}  \bar s_{H} C_{R}(z; \lambda_{n}, c) \cr  \bar c_{H} \check S_{R} (z; \lambda_{n}, c)  \end{pmatrix} ,  \cr
\noalign{\kern 5pt}
&\bar {\bf f}_{Rn}^c (z)  = \frac{1}{\sqrt{r^c_n}} 
\begin{pmatrix}  \bar s_{H} \hat C_{R}(z; \lambda_{n}, c) \cr  \bar c_{H}  S_{R} (z; \lambda_{n}, c)  \end{pmatrix} , ~~
\bar {\bf f}_{Rn}^d (z)  = \frac{1}{\sqrt{r^d_n}} 
\begin{pmatrix}  - \bar c_{H} \check C_{R}(z; \lambda_{n}, c) \cr  \bar s_{H}  S_{R} (z; \lambda_{n}, c)  \end{pmatrix} ,\cr
\noalign{\kern 5pt}
&\bar {\bf f}_{Ln}^a (z)  = \frac{1}{\sqrt{r^a_n}} 
\begin{pmatrix} \bar s_{H} \hat S_{L}(z; \lambda_{n}, c) \cr \bar c_{H} C_{L} (z; \lambda_{n}, c) \end{pmatrix} , ~~
\bar {\bf f}_{Ln}^b (z)  = \frac{1}{\sqrt{r^b_n}} 
\begin{pmatrix} - \bar c_{H} \check S_{L}(z; \lambda_{n}, c) \cr \bar s_{H} C_{L} (z; \lambda_{n}, c) \end{pmatrix} , \cr
\noalign{\kern 5pt}
&\bar {\bf f}_{Ln}^c (z)  = \frac{1}{\sqrt{r^c_n}} 
\begin{pmatrix} \bar c_{H} S_{L}(z; \lambda_{n}, c) \cr  - \bar s_{H} \hat C_{L} (z; \lambda_{n}, c) \end{pmatrix} , ~~
\bar {\bf f}_{Ln}^d (z)  = \frac{1}{\sqrt{r^d_n}} 
\begin{pmatrix}  \bar s_{H}  S_{L}(z; \lambda_{n}, c) \cr \bar c_{H} \check C_{L} (z; \lambda_{n}, c) \end{pmatrix} , \cr
\noalign{\kern 5pt}
&\qquad \bar c_H = \cos \onehalf \theta_H ~,~ \bar s_H = \sin \onehalf \theta_H ~, \cr
&\qquad  r_n = \int_1^{z_L} dz  \big\{ | \hat f_{n} (z) |^{2} + | \hat g_{n} (z) |^{2} \big\} 
\quad {\rm for}~ \begin{pmatrix} \hat f_{n} (z) \cr \hat g_{n} (z) \end{pmatrix} .
\label{RSfermionKKB2}
\end{align}
Functions $\hat S_{R/L}, \check S_{R/L}$ etc. are defined in (\ref{functionA4}).
In (\ref{RSfermionKKB1}) two expressions in an overlapping region in $\theta_H$ are the same.

\subsection{Fermion fields $\chi_{R/L}^{(n)}$ for $c=0$}

For $c=0$ $C_{R/L}(z; \lambda, 0)$ and $S_{R/L}(z; \lambda, 0)$ reduce to trigonometric functions.
\begin{align}
\begin{pmatrix} C_L \cr S_L \end{pmatrix} (z; \lambda, 0) &= 
\begin{pmatrix} \cos \lambda (z - z_L) \cr  \sin \lambda (z - z_L) \end{pmatrix} , \cr
\noalign{\kern 5pt}
\begin{pmatrix} C_R \cr S_R \end{pmatrix} (z; \lambda, 0) &= 
\begin{pmatrix} \cos \lambda (z - z_L) \cr  - \sin \lambda (z - z_L) \end{pmatrix} .
\label{functionB1}
\end{align}
The spectrum and wave functions in $1 \le z = e^{ky}  \le z_L$ in the original gauge are given
for  type 1A  by
\begin{align}
&\underline{\hbox{type 1A}} : \cr
\noalign{\kern 5pt}
&\lambda_n = \frac{1}{z_L -1}  \big| n \pi + \onehalf \theta_H \big| \quad (- \infty < n < \infty), \cr
\noalign{\kern 5pt}
&\begin{pmatrix} f_{Rn} (y) \cr \noalign{\kern 5pt} g_{Rn} (y) \end{pmatrix} = \frac{1}{\sqrt{z_L -1}}
\begin{pmatrix} \cos \Big\{ (n \pi + \onehalf \theta_H) \myfrac{z - z_L}{z_L -1} + \onehalf \theta(z) \Big\} \cr
\sin \Big\{ (n \pi + \onehalf \theta_H) \myfrac{z - z_L}{z_L -1} + \onehalf \theta(z) \Big\}  \end{pmatrix} , \cr
\noalign{\kern 5pt}
&\begin{pmatrix} f_{Ln} (y) \cr \noalign{\kern 5pt} g_{Ln} (y) \end{pmatrix} = \frac{1}{\sqrt{z_L -1}}
\begin{pmatrix} - \sin \Big\{ (n \pi + \onehalf \theta_H) \myfrac{z - z_L}{z_L -1} + \onehalf \theta(z) \Big\} \cr
\cos \Big\{ (n \pi + \onehalf \theta_H) \myfrac{z - z_L}{z_L -1} + \onehalf \theta(z) \Big\}  \end{pmatrix} , 
\label{RSfermionKKB3}
\end{align}
and for type 2A by
\begin{align}
&\underline{\hbox{type 2A}} : \cr
\noalign{\kern 5pt}
&\lambda_n = \frac{1}{z_L -1}  \big|( n+ \onehalf)  \pi +  \onehalf \theta_H \big| \quad (- \infty < n < \infty), \cr
\noalign{\kern 5pt}
&\begin{pmatrix} f_{Rn} (y) \cr \noalign{\kern 5pt} g_{Rn} (y) \end{pmatrix} = \frac{1}{\sqrt{z_L -1}}
\begin{pmatrix}  - \sin \Big\{ (n \pi + \onehalf \pi + \onehalf \theta_H) \myfrac{z - z_L}{z_L -1} + \onehalf \theta(z) \Big\} \cr
\cos \Big\{ (n \pi + \onehalf \pi+ \onehalf \theta_H) \myfrac{z - z_L}{z_L -1} + \onehalf \theta(z) \Big\}  \end{pmatrix} , \cr
\noalign{\kern 5pt}
&\begin{pmatrix} f_{Ln} (y) \cr \noalign{\kern 5pt} g_{Ln} (y) \end{pmatrix} = \frac{1}{\sqrt{z_L -1}}
\begin{pmatrix}  \cos \Big\{ (n \pi + \onehalf \pi+ \onehalf \theta_H) \myfrac{z - z_L}{z_L -1} + \onehalf \theta(z) \Big\} \cr
\sin \Big\{ (n \pi + \onehalf \pi+ \onehalf \theta_H) \myfrac{z - z_L}{z_L -1} + \onehalf \theta(z) \Big\}  \end{pmatrix} .
\label{RSfermionKKB4}
\end{align}
Note that the expressions (\ref{RSfermionKKB3}) and (\ref{RSfermionKKB4}) reduce, up to normalization factors, 
 to the expressions (\ref{fermionKKflat2}) and (\ref{fermionKKflat3}) in the flat spacetime limit, respectively.
For other regions in $y$, the wave functions are defined by (\ref{RSfermionKK3}).

\vskip 1.cm

\def\jnl#1#2#3#4{{#1}{\bf #2},  #3 (#4)}

\def\Zphys{{\em Z.\ Phys.} }
\def\jssc{{\em J.\ Solid State Chem.\ }}
\def\jpsJ{{\em J.\ Phys.\ Soc.\ Japan }}
\def\ptps{{\em Prog.\ Theoret.\ Phys.\ Suppl.\ }}
\def\PTP{{\em Prog.\ Theoret.\ Phys.\  }}
\def\PTEP{{\em Prog.\ Theoret.\ Exp.\  Phys.\  }}
\def\JMP{{\em J. Math.\ Phys.} }
\def\NPB{{\em Nucl.\ Phys.} B}
\def\NP{{\em Nucl.\ Phys.} }
\def\PLB{{\it Phys.\ Lett.} B}
\def\PL{{\em Phys.\ Lett.} }
\def\PRL{\em Phys.\ Rev.\ Lett. }
\def\PRB{{\em Phys.\ Rev.} B}
\def\PRD{{\em Phys.\ Rev.} D}
\def\PRe{{\em Phys.\ Rep.} }
\def\AP{{\em Ann.\ Phys.\ (N.Y.)} }
\def\RMP{{\em Rev.\ Mod.\ Phys.} }
\def\ZPC{{\em Z.\ Phys.} C}
\def\SCI{\em Science}
\def\CMP{\em Comm.\ Math.\ Phys. }
\def\MPLA{{\em Mod.\ Phys.\ Lett.} A}
\def\IJMPA{{\em Int.\ J.\ Mod.\ Phys.} A}
\def\IJMPB{{\em Int.\ J.\ Mod.\ Phys.} B}
\def\EPJC{{\em Eur.\ Phys.\ J.} C}
\def\PR{{\em Phys.\ Rev.} }
\def\JHEP{{\em JHEP} }
\def\JCAP{{\em JCAP} }
\def\cmp{{\em Com.\ Math.\ Phys.}}
\def\JPA{{\em J.\  Phys.} A}
\def\JPG{{\em J.\  Phys.} G}
\def\NJP{{\em New.\ J.\  Phys.} }
\def\CQG{\em Class.\ Quant.\ Grav. }
\def\ATMP{{\em Adv.\ Theoret.\ Math.\ Phys.} }
\def\ibid{{\em ibid.} }
\def\ChP{{\em Chin.Phys.}C}
\def\NCA{{\it Nuovo Cim.} A}


\renewenvironment{thebibliography}[1]
         {\begin{list}{[$\,$\arabic{enumi}$\,$]}  
         {\usecounter{enumi}\setlength{\parsep}{0pt}
          \setlength{\itemsep}{0pt}  \renewcommand{\baselinestretch}{1.2}
          \settowidth
         {\labelwidth}{#1 ~ ~}\sloppy}}{\end{list}}

\vskip 1.cm

\leftline{\Large \bf References}


\end{document}